\documentclass[twocolumn,aps,pra,superscriptaddress,longbibliography]{revtex4-2}
\usepackage{amsmath}
\usepackage{enumitem}
\usepackage{amssymb}
\usepackage{adjustbox}
\usepackage{bm}
\usepackage{hyperref}
\hypersetup{
	colorlinks = true,
	linkcolor = blue,
	citecolor = blue,
	urlcolor  = blue,
}
\usepackage[dvipsnames]{xcolor}
\usepackage[utf8]{inputenc}
\usepackage{graphicx}
\usepackage{dcolumn}
\usepackage{bm}
\usepackage{physics}
\usepackage{wrapfig}
\usepackage{xcolor}
\usepackage[normalem]{ulem}
\newcommand{\stkout}[1]{\ifmmode\text{\sout{\ensuremath{#1}}}\else\sout{#1}\fi}

\begin{document}

\title{Why is topology hard to learn?}

\date{\today}
\begin{abstract}
Much attention has been devoted to the use of machine learning to approximate physical concepts. Yet, due to challenges in interpretability of machine learning techniques, the question of what physics machine learning models are able to learn remains open. Here we bridge the concept a physical quantity and its machine learning approximation in the context of the original application of neural networks in physics: topological phase classification. We construct a hybrid tensor-neural network object that exactly expresses real space topological invariant and rigorously assess its trainability and generalization. Specifically, we benchmark the accuracy and trainability of a tensor-neural network to multiple types of neural networks, thus exemplifying the differences in trainability and representational power. Our work highlights the challenges in learning topological invariants and constitutes a stepping stone towards more accurate and better generalizable machine learning representations in condensed matter physics.
\end{abstract}

\author{D. O. Oriekhov}
\email{d.oriekhov@tudelft.nl}
\affiliation{QuTech and Kavli Institute of Nanoscience, Delft University of Technology, 2628 CJ Delft, the Netherlands}
\author{Stan Bergkamp}
\affiliation{QuTech and Kavli Institute of Nanoscience, Delft University of Technology, 2628 CJ Delft, the Netherlands}
\author{Guliuxin Jin}
\affiliation{QuTech and Kavli Institute of Nanoscience, Delft University of Technology, 2628 CJ Delft, the Netherlands}%
\author{Juan Daniel Torres Luna}
\affiliation{QuTech and Kavli Institute of Nanoscience, Delft University of Technology, 2628 CJ Delft, the Netherlands}%
\author{Badr Zouggari}
\affiliation{QuTech and Kavli Institute of Nanoscience, Delft University of Technology, 2628 CJ Delft, the Netherlands}%
\author{Sibren van der Meer}
\affiliation{QuTech and Kavli Institute of Nanoscience, Delft University of Technology, 2628 CJ Delft, the Netherlands}%
\author{Naoual El Yazidi}
\affiliation{QuTech and Kavli Institute of Nanoscience, Delft University of Technology, 2628 CJ Delft, the Netherlands}%
\author{Eliska Greplova}%
\email{e.greplova@tudelft.nl}
\affiliation{QuTech and Kavli Institute of Nanoscience, Delft University of Technology, 2628 CJ Delft, the Netherlands}%

\maketitle


\emph{Introduction.}
Phase classification has become a prototypical benchmark for data-driven analysis of condensed matter physics. The type and complexity of the phase transition dictate the level of complexity of the algorithm one has to employ. This topic has been broadly explored, offering a menu of both supervised and unsupervised techniques ranging from simple clustering \cite{Wang2016_DiscoveringPT,HuSinghScalettar2017_UnsupervisedCrit,YueWangLyu2022_IncrementalIsingPCA} to more complex machine learning methods \cite{Carrasquilla_Melko_2017,DengLiDasSarma2017_MLTopStates,ZhangShenZhai2018_TopInvNN, vNLH-confusion}. The phase classification problem is most commonly posed like so: we allow our model to view a dataset that is both relevant and straightforwardly obtainable in the scenario we wish to study. We introduce this data set to a model that has no prior knowledge of underlying physics. Then we proceed to optimize the model to determine which of the available phases of matter a given sample in the dataset corresponds to.

Within data-driven phase classification, the topological phases are of particular interest \cite{Zhang2020PhysRevResearch.2.023283,ZhangShenZhai2018_TopInvNN}. The presence of topology in the condensed matter systems is a field of an active research interest \cite{mondragon2014topological,Chiu2016RevModPhys.88.035005,HasanKane2010,QiZhang2011,BernevigHughesZhang2006,Fu2011,Zhang2009,YanZhang2012,Rachel2018,AndoFu2015}. In particular, the formulation of appropriate topological invariants is a highly non-trivial complex venture. When it comes to experimental verification \cite{de2019observation,splitthoff2024gate,Kanungo_2022,kiczynski2022engineering,jouanny2024band,mei2018robust,zheng2022engineering,kim2021quantum,vega2021qubit} of the theoretical predictions, there is an additional hurdle: the majority of topological invariants are formulated in the k-space \cite{AltlandZirnbauer1997_AZ, SchnyderRyuFurusakiLudwig2008,Kitaev2009_PeriodicTable,DengLiDasSarma2017_MLTopStates} and their real-space counterparts do not necessarily produce the same phase diagram in a finite size setting~\cite{mondragon2014topological,jin2024topologicalfinitesizeeffect}.

Approaching the classification of topological phases from the data-driven perspective, we arrive at a seemingly contradictory observation: topological invariants are inherently global features, but even simple machine learning can immediately distinguish states with and without topological constraints in a both supervised \cite{Carrasquilla_Melko_2017} and unsupervised \cite{Greplova_Valenti_Boschung_Huber_2020} way. Machine learning models have been shown to predict Chern numbers \cite{Sun2018PhysRevB.98.085402,Zhang2017PhysRevLett.118.216401,Baireuther2023}, winding numbers \cite{ZhangShenZhai2018_TopInvNN,caio2019machine,Zhang2021nonHerm}, and classify the number of topological phases \cite{Greplova_Valenti_Boschung_Huber_2020,Molignini2021PostPhys.11.3.073,Ghosh2024PRB,Ghosh2025IOP}. 

We conclude that, on the one hand, it is an exceedingly difficult venture to identify the presence of topology by traditional methods that rely on formulation and evaluation of a global invariant. On the other hand, even a simple neural network can evidently learn to predict the values of these invariants, even from limited data. Are the machine learning models somehow so powerful as to completely bypass the complexity of topological invariants and reliably approximate it from a labeled dataset?

In this work, we set out to rigorously resolve this contradiction and bridge the condensed matter approach of constructing the global invariants with the data-driven machine learning approaches. We do so in three steps: \emph{Step 1:} We construct a hybrid tensor-neural network that \emph{exactly} expresses a known topological invariant: real space winding number for a general AIII symmetry class system. \emph{Step 2:} We benchmark the performance and generalization of this model for the (extended) Su-Schriefer-Heeger model against different variations of physics-agnostic neural networks. \emph{Step 3:} We analyze the behavior of weights during the training and relate it to the original topological invariant formula.

\begin{figure*}
		\centering
        \includegraphics[scale=1.1]{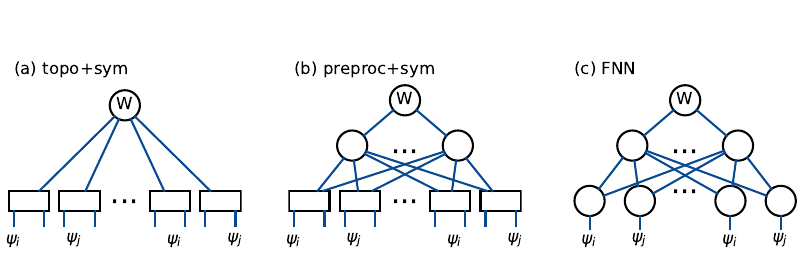}
        \caption{Neural network architectures: (a) \emph{topological} neural network that combines a tensor layer and one neural layer that exactly reproduces the structure of real space winding number for SSH-like two-band models within the AIII symmetry class. 
        (b) \emph{preprocessed} neural network: the tensor layer remains included, but the remainder of the network consists of multi-layer classifier that no longer exactly corresponds to the topological invariant.
        (c) \emph{feed-forward} neural network is a fully-connected  classifier without the preprocessing layer and without the default correspondence to the topological invariant. 
        Input size for (a) and (b) networks scales as $N^2$ for $N$ unit cells. The input size for network (c) is $4N^2$.}
		\label{fig:1}
	\end{figure*}

We show that it is indeed possible to construct a neural network model that learns the topological invariant \emph{exactly} and not simply a proxy within the dataset that allows for effective minimization of the loss function but has no physics meaning. We find that our exact topological neural network approximates the topological invariant with four orders of magnitude better precision than a traditional feed-forward neural networks, when trained. However, this achievement comes with a price: by expressing the topological invariant exactly, we leave the overparametrized regime in which neural networks thrive, and the model becomes increasingly sensitive to initialization. Below, we build a side-by-side comparison of designing and training the exact topological model and its physics agnostic approximation, and specifically highlight the connection between the real space winding number formula and a neural network.

\emph{Results.} 
The topological invariant we choose to approximate is the real space winding number (RSWN), originally derived for the general AIII class of topological insulators in Ref.~\cite{mondragon2014topological}. For a one-dimensional system with the chiral symmetry operator $\Gamma$, the RSWN reads:
\begin{align}
\label{eq:RSWN}
W=-\frac{1}{L} \operatorname{Tr}\left\{P_B Q P_A\left[X, P_A Q P_B\right]\right\}.
\end{align}
It is expressed through the commutator of $Q$-operator, and projectors $P_{A,B}$ on two sets of sublattices $A,B$ conjugated by the $\Gamma=P_A-P_B$ operator, with the coordinate operator $X$. The operator $Q=P_{+}-P_{-}$ is composed of projectors on empty and filled bands, and the system length $L$ equals the number of unit cells $N$ times the lattice constant. In this work we focus on the winding number for half-filling; thus, the operators $P_{\pm}$ correspond to positive/negative energy bands. 
Substituting the components of eigenstates, one finds that the formula \eqref{eq:RSWN} translates into a polynomial of the fourth order (see Sec.S1 in Supplemental material \cite{Supplement}).

The prototypical example of the AIII topological insulator is the Su-Schriefer-Heeger (SSH) family of Hamiltonians~\cite{su1979solitons,HasanKane2010,asboth2016short}. These models have two sublattices, exhibit time-reversal symmetry, and their Hamiltonian has the form:
\begin{align}
\begin{aligned}
& H=\sum_{n=1}^N\left(v c_{n, A}^{\dagger} c_{n, B}+w c_{n, B}^{\dagger} c_{n+1, A}\right. \\
&\left.+J_3 c_{n, B}^{\dagger} c_{n+2, A}+J_3^{\prime} c_{n, A}^{\dagger} c_{n+1, B}\right)+ \text { h.c., }
\end{aligned}
\label{eq:ssh}
\end{align}
where $c_{n, \alpha}^{\dagger}\left(c_{n, \alpha}\right)$ denotes the ladder operator of a quasiparticle at lattice site $(n, \alpha)$, with the unit cell index $n \in[1, N]$ and the sublattice index $\alpha \in A, B$. The ratio of intra-cell to inter-cell hopping, $v / w$, controls the topological phase transition for SSH model. The hopping amplitudes $J_3$ and $J_3'$ correspond to the next-nearest neighbor hopping mechanism, and in their presence, the model is referred to as an extended SSH model. The phase diagram of the SSH model with an even number of sites contains two phases with winding numbers $1$ and $0$. These phases correspond to $w<v$ and $w>v$ relations of hopping parameters in the infinite chain limit. For the finite-size chain, the hopping ratio corresponding to the phase transition point becomes size-dependent. The phase diagram of the extended SSH model contains winding numbers from $-1$ to $2$ \cite{asboth2016short,HasanKane2010,su1979solitons,perez2018ssh,jin2024topologicalfinitesizeeffect}.

For the family of Hamiltonians described by Eq.~\eqref{eq:ssh}, the RSWN formula in Eq.~\eqref{eq:RSWN} can be reduced to the following expression
\begin{align}
\label{eq:reduced-rswn}
	W&=2 \sum_{n\notin\{\text{edge st.}\}} \sum_{x}x \left(\left[\Psi_{x, A}^{n,+}\right]^2-\left[\Psi_{x, B}^{n,+}\right]^2 \right)+\nonumber\\
    &\begin{cases}
		2 \sum_{n \in\{\text{hybr. edge states}\}} \sum_{x}x \left(\left[\Psi_{x, A}^{n,+}\right]^2-\left[\Psi_{x, B}^{n,+}\right]^2 \right), 
        \\
		\begin{array}{c}
        4 \sum\limits_{n \in\{\text{loc. edge st., B}\}} \sum_{x}x \left[\Psi_{x, A}^{n,+}\right]^2-\\
        -4 \sum\limits_{n \in\{\text{loc. edge st., A}\}} \sum_{x}x\left[\Psi_{x, B}^{n,+}\right]^2.
        \end{array}
	\end{cases}
\end{align}
Here, the summation goes over positive energy band eigenstate indices $n$ with energy $E_n$ and coordinates of the unit cells $x$. The $\Psi_{x,A/B}^{n,+}$ is a component of the eigenstate $n$ with $E_n>0$ at the unit cell with coordinate $x$ and site $A/B$. The summation includes differences of wave functions on the different sublattices $A$ and $B$ within the same unit cells. 
The derivation of this formula is shown in detail in Sec.S1 in \cite{Supplement}. Because of the finite-size effect on edge state localization properties  \cite{jin2024topologicalfinitesizeeffect}, the formula contains two types of contributions. First type corresponds to hybridized edge states, and second - to the single-side localized states, distinguished by the corresponding edge of chain and sublattice family of the main wave function weight $A / B$. 

Notable aspect of expression in Eq.~\eqref{eq:reduced-rswn}, that even though it is a rather complicated summation, it is a linear combination of the products of pairs of the SSH wave-function amplitudes. Eq.~\eqref{eq:reduced-rswn} can thus immediately inform our neural network construction for RSWN. Our architecture is shown in Fig.~\ref{fig:1}a. We first pre-process the wave-function amplitudes via a tensor layer that forms the two-term product pairs. These are then connected, via a single layer of weights, to the output neuron that stores the value of RSWN, $W$. From now on, we refer to this model as \emph{topological}, since for specific fixed values of its weights (blue lines in Fig.~\ref{fig:1}a), the formula~\eqref{eq:reduced-rswn} is expressed exactly. In Fig.~\ref{fig:1}b, we show a natural generalization of the topological model: we maintain the tensor layer preprocessing and connect it to a two-layer generic neural network with ReLU activation functions. This network no longer expresses RSWN exactly, but takes correctly paired wavefunction amplitudes as its inputs and has a larger space of trainable parameters to approximate RSWN. We refer to the network as \emph{preprocessed}. Finally, in Fig.~\ref{fig:1}c we show a fully connected feed-forward network (FNN) that takes the wavefunction amplitudes as an input individually and is trained to approximate RSWN (for architecture details see Sec.5 of \cite{Supplement}).

\begin{figure}
		\centering
		\includegraphics[scale=1.0]{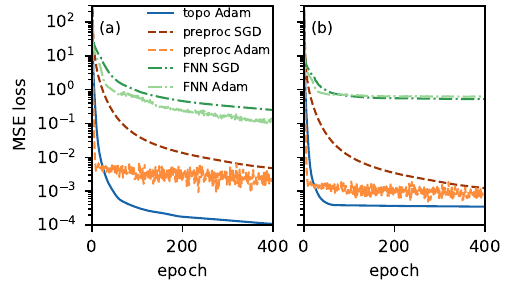}
		\caption{MSE loss for the RSWN regression: Panel (a) shows MSE loss on training data, and (b) on validation. Three models were used to learn the float value of RSWN. The topological model (`topo') is in blue, the preprocessed model (`pre-proc') is in red and orange for SGD and Adam, respectively. FNN model is in dark (light) green for SGD (Adam).}
		\label{fig:2}
	\end{figure}

In Fig.~\ref{fig:2} we show the loss function, mean-squared error (MSE) with respect to the correct value of RSWN, as a function of training epoch. We train on the dataset of approximately 140k samples. Each sample is a 2D array of wavefunction components ordered as a function of lattice site and eigenstate index.
Each sample is generated from a Hamiltonian in Eq.~\eqref{eq:ssh} with randomly chosen values of hopping amplitudes. The generated set of Hamiltonians has approximately equal distributions over different winding number classes. Half of the dataset is generated using the SSH Hamiltonian, and the other half is generated using the ESSH Hamiltonian.

We show the loss of the topological model in blue, the preprocessed model in red (trained with stochastic gradient descent (SGD)) and in orange (trained with Adam optimizer), and FFN in dark green (SGD) and light green (Adam). The panel (a) shows training, and panel (b) shows validation. We see that the topological model performs by far the best and shows the fastest convergence. The initialization of the topological model was performed in a deterministic way: first, we take the set of weights that match exactly the RSWN formula in Eq.~\eqref{eq:reduced-rswn}, where $w_i=1$ for all $i$. Secondly, we randomize the weights by adding a random tensor generated from a Gaussian distribution (with mean $0.4$ and variance $0.15$) with a fixed seed for reproducibility. Then we begin to train. We will analyze the robustness of the convergence and its relation to initialization in the next paragraphs.

\begin{figure}
		\centering
        \includegraphics[scale=1]{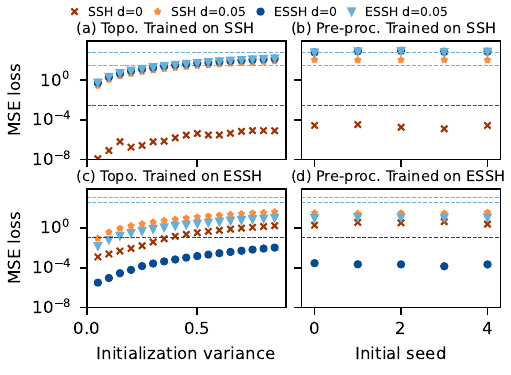}
		\caption{
        Generalization of the topological, pre-processed and FNN models: each neural network is trained on one non-disordered $d=0$ type of dataset corresponding to a model Hamiltonian of SSH (a, b) or ESSH (c, d), and tested on another Hamiltonian dataset as well as both disordered types (color is indicated in legend).
        Model types are indicated in the titles of panels.
        We train a topological model (a,c) for different initialization variances and a pre-processed model (b, d) for five different seeds.
        The FNN model results are the dashed lines in each panel.
        }
		\label{fig:3}
	\end{figure}
    \begin{figure*}
		\centering
		\includegraphics[scale=1]{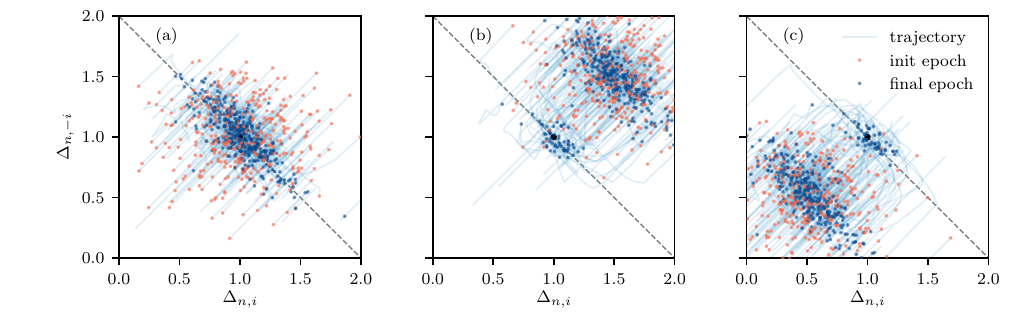}
		\caption{An evolution of the topological model weights during training: the orange dots are initial values of each weight, the blue line corresponds to the training trajectory and dark blue dot is the final point. In panel (a) the model initialization is centered around the correct RSWN formula, the weights are initialized from the Gaussian distribution centered around $1.0$; (b) the Gaussian distribution for weights initialization is shifted away from the correct RSWN formula and is centered around $1.5$; (c) the Gaussian distribution for the weights initialization is centered around $0.5$.
        }
		\label{fig:4}
\end{figure*}

The next critical aspect we need to investigate is the generalization of our model. In principle, it is simple to construct a model that has a very low loss on a given dataset. The challenge is to make the model generalize correctly. This aspect is particularly important when one wishes to use data-driven techniques to reliably determine topology: a key feature of topological invariants is that they robustly predict topological phase even in the presence of disorder. In Sec. S4 of \cite{Supplement} we explain in depth a simple example of a poor generalization: we show that most of the amplitudes of any eigenstate of SSH models are monotonous functions of the ratio $v/w$ which determines the phase transition. As such, an example of a model that can predict whether a given state is in the trivial or topological phase is a value of almost any single wave function amplitude. We show, that, unsurprisingly such a model is not resilient against disorder and becomes invalid almost immediately. The in-depth study of generalization of physics agnostic neural networks on the SSH Hamiltonian was shown in Ref.~\cite{Cybinski2024}.

In Fig.~\ref{fig:3}, we show an analysis of the generalization for all three types of models in Fig.~\ref{fig:1}, between SSH and ESSH models and in the presence of disorder. The red color family denotes SSH, blue color family ESSH. The dashed lines show the values of FFN neural network. In Fig.~\ref{fig:3}(a) we show MSE loss as a function of the initialization variance for the topological model trained on SSH half of our dataset. Here, we follow the initialization strategy explained above, but vary the mean of the Gaussian distribution from which we sample the weights from. Unsurprisingly, SSH at zero disorder ($d=\delta t/t=0$) exhibits very low MSE values across the initialization range with slight increase as a function of the initialization variance. This increase is upper-bounded by the MSE value of the physics agnostic FFN trained on this data set. Disordered data for SSH and both disordered and non-disordered data for ESSH behave similarly, albeit with bigger absolute value, and are upper-bounded by their respective FNN. The relative disorder $d$ of hopping parameters $v,w$ in SSH model, and $J_3,J_3'$ in ESSH model was equal to $0.05$ in all data samples. In Fig.~\ref{fig:3}(b,c,d) we perform analogous analysis. We train the pre-processed model on SSH, topological model on ESSH and the pre-processed model on ESSH, respectively. 
We observe that topological models have better generalization properties, especially at lower initialization variances. Their error slowly increases until it reaches roughly the value of the pre-processed network, which is in turn always upper-bounded by values of FNN.

A common way to gain insight into the interpretability of a machine learning model is to analyze its weight space~\cite{Tsang2017_InteractionsWeights,Eilertsen2020_ClassifyingClassifier,hernandes2025adiabatic}. In the instance of the topological neural network and its corresponding formula in Eq.~\eqref{eq:reduced-rswn}, the weight space analysis can be particularly interpretable. Firstly, the architecture itself is simple and after tensor pre-processing, it only consists of a single trainable layer. Specifically, the kernel of the tensor is multiplied with the weights of the single layer, $\Delta_{n,i}$, and a bias is added. This hybrid tensor-neural network then outputs the topological invariant, $W$. More, importantly, the symmetry properties of Eq.~\eqref{eq:reduced-rswn} dictate that all the weights need to lie on the line $\Delta_{n,x}+\Delta_{n,-x}=2$, where index $n$ corresponds to the eigenstate and index $x$ to the site of the chain (see Sec. S1 of \cite{Supplement} for details). In Fig.~\ref{fig:4}, we visualize the weight trajectories during the training for three different initializations (panels (a), (b), (c) respectively). The orange dots specify the initialization point for each weight, the blue dots the final point during training and the light blue line the trajectory. Each panel corresponds to the different instance of Gaussian initialization initialized with a mean of $1.0$, $1.5$, $0.5$, respectively. All distributions have a variance of $0.3$.
For this analysis, we added disorder to our initial dataset. Specifically, we introduce an inversion symmetric disorder to the parameters $w, v, J$. Disorder strength, $d = \delta \{w,v,J\}/\{w,v,J\}$ was chosen to be $d=0.05$ for $75\%$ of the samples and $d=0.08$ for $25\%$ of the samples. We constructed this disorder explicitly to violate symmetries in the dataset, such that we can visualize the weights converging on a simple line instead of a more complicated pattern. Additionally, we combined two disorder patterns to prevent the model from fixating on specific amplitudes within the dataset with a uniform disorder (see Secs. S2 and S3 in Ref.~\cite{Supplement}).

In Fig.~\ref{fig:4}a we see that the weights converge towards the required expression, while in the panels (b) and (c) only a portion of the weights converges on the correct line, while the rest of the weights converge to another line-like structure, albeit with the constant offset to the actual symmetry line. This means that the network converged to a suboptimal local minimum,  but that it also found a `new' RSWN formula with a similar structure in the process.

\emph{Conclusions and Outlook.}
In this work, we set out to answer the question \emph{`Why is topology hard to learn?'}. The reason for this difficulty is that trained machine learning models most commonly learn a dataset dependent proxy rather than the topological invariant itself. Here we have shown how to construct a trainable object that learns the topological quantity exactly. However, constructing such precise topological invariant representation comes hand-in-hand with hindrances in trainability. While the topological network we construct is less trainable, it exemplifies the difference between learning the physical quantity itself and learning its machine learning approximations. Formulating this distinction precisely opens up an avenue to the more informed and interpretable application of machine learning models in condensed matter physics.

Going forward we could apply the approach presented here to enhance interpretability for the systems with unknown topological order parameters. One could combine the tensor network layer technique with the multi-kernel approach from convolutional neural networks. In particular, combining several tensor network layers of different rank and performing bottleneck procedure \cite{Cybinski2024Tetris}, one could extract the main structure of topological invariant out of several candidates. There is also exciting possible connection with Kolmogorov-Arnold networks (KANs) \cite{liu2025kankolmogorovarnoldnetworks}. The key ingredient of the topological networks presented in this work is the tensor transformation layer. This layer converts a non-linear polynomial structure of topological invariant into a linear task, well-suited for linear perceptron neural nets. Kolmogorov-Arnold networks have been shown to be well-suited for approximating polynomial functions. In this regard we point out that tensor-neural construction of our topological network could be viewed as a bridge between KANs and linear perceptron. Namely, a first tensor layer could be composed out of several KAN layers described in~Ref.~\cite{liu2025kankolmogorovarnoldnetworks}. 

\begin{acknowledgements}
We are thankful for fruitful discussions with Ana Silva, Thomas Spriggs, Anton Akhmerov, Mohammed Boky, and Anna Dawid. This work is part of the project Engineered Topological Quantum
 Networks (Project No.VI.Veni.212.278) of the research
 program NWO Talent Programme Veni Science domain
 2021 which is financed by the Dutch Research Council (NWO). D.O.O. acknowledges the support by the Kavli Foundation. GJ acknowledges the research
 program ``Materials for the Quantum Age'' (QuMat) for
 financial support.
\end{acknowledgements}

\begin{center}
    {\bf Data availability statement}
\end{center}
The code to reproduce research in this paper can be found at \cite{Supplement-code}.

\bibliography{topo_learn_ref_with_doi.bib}  

\begin{thebibliography}{53}%
\makeatletter
\providecommand \@ifxundefined [1]{%
 \@ifx{#1\undefined}
}%
\providecommand \@ifnum [1]{%
 \ifnum #1\expandafter \@firstoftwo
 \else \expandafter \@secondoftwo
 \fi
}%
\providecommand \@ifx [1]{%
 \ifx #1\expandafter \@firstoftwo
 \else \expandafter \@secondoftwo
 \fi
}%
\providecommand \natexlab [1]{#1}%
\providecommand \enquote  [1]{``#1''}%
\providecommand \bibnamefont  [1]{#1}%
\providecommand \bibfnamefont [1]{#1}%
\providecommand \citenamefont [1]{#1}%
\providecommand \href@noop [0]{\@secondoftwo}%
\providecommand \href [0]{\begingroup \@sanitize@url \@href}%
\providecommand \@href[1]{\@@startlink{#1}\@@href}%
\providecommand \@@href[1]{\endgroup#1\@@endlink}%
\providecommand \@sanitize@url [0]{\catcode `\\12\catcode `\$12\catcode
  `\&12\catcode `\#12\catcode `\^12\catcode `\_12\catcode `\%12\relax}%
\providecommand \@@startlink[1]{}%
\providecommand \@@endlink[0]{}%
\providecommand \url  [0]{\begingroup\@sanitize@url \@url }%
\providecommand \@url [1]{\endgroup\@href {#1}{\urlprefix }}%
\providecommand \urlprefix  [0]{URL }%
\providecommand \Eprint [0]{\href }%
\providecommand \doibase [0]{https://doi.org/}%
\providecommand \selectlanguage [0]{\@gobble}%
\providecommand \bibinfo  [0]{\@secondoftwo}%
\providecommand \bibfield  [0]{\@secondoftwo}%
\providecommand \translation [1]{[#1]}%
\providecommand \BibitemOpen [0]{}%
\providecommand \bibitemStop [0]{}%
\providecommand \bibitemNoStop [0]{.\EOS\space}%
\providecommand \EOS [0]{\spacefactor3000\relax}%
\providecommand \BibitemShut  [1]{\csname bibitem#1\endcsname}%
\let\auto@bib@innerbib\@empty
\bibitem [{\citenamefont {Wang}(2016)}]{Wang2016_DiscoveringPT}%
  \BibitemOpen
  \bibfield  {author} {\bibinfo {author} {\bibfnamefont {L.}~\bibnamefont
  {Wang}},\ }\bibfield  {title} {\bibinfo {title} {Discovering phase
  transitions with unsupervised learning},\ }\href
  {https://doi.org/10.1103/PhysRevB.94.195105} {\bibfield  {journal} {\bibinfo
  {journal} {Physical Review B}\ }\textbf {\bibinfo {volume} {94}},\ \bibinfo
  {pages} {195105} (\bibinfo {year} {2016})}\BibitemShut {NoStop}%
\bibitem [{\citenamefont {Hu}\ \emph {et~al.}(2017)\citenamefont {Hu},
  \citenamefont {Singh},\ and\ \citenamefont
  {Scalettar}}]{HuSinghScalettar2017_UnsupervisedCrit}%
  \BibitemOpen
  \bibfield  {author} {\bibinfo {author} {\bibfnamefont {W.}~\bibnamefont
  {Hu}}, \bibinfo {author} {\bibfnamefont {R.~R.~P.}\ \bibnamefont {Singh}},\
  and\ \bibinfo {author} {\bibfnamefont {R.~T.}\ \bibnamefont {Scalettar}},\
  }\bibfield  {title} {\bibinfo {title} {Discovering phases, phase transitions
  and crossovers through unsupervised machine learning: A critical
  examination},\ }\href {https://doi.org/10.1103/PhysRevE.95.062122} {\bibfield
   {journal} {\bibinfo  {journal} {Physical Review E}\ }\textbf {\bibinfo
  {volume} {95}},\ \bibinfo {pages} {062122} (\bibinfo {year}
  {2017})}\BibitemShut {NoStop}%
\bibitem [{\citenamefont {Yue}\ \emph {et~al.}(2022)\citenamefont {Yue},
  \citenamefont {Wang},\ and\ \citenamefont
  {Lyu}}]{YueWangLyu2022_IncrementalIsingPCA}%
  \BibitemOpen
  \bibfield  {author} {\bibinfo {author} {\bibfnamefont {Z.}~\bibnamefont
  {Yue}}, \bibinfo {author} {\bibfnamefont {Y.}~\bibnamefont {Wang}},\ and\
  \bibinfo {author} {\bibfnamefont {P.}~\bibnamefont {Lyu}},\ }\bibfield
  {title} {\bibinfo {title} {Incremental learning of phase transition in ising
  model: Preprocessing, finite-size scaling and critical exponents},\ }\href
  {https://doi.org/10.1016/j.physa.2022.127538} {\bibfield  {journal} {\bibinfo
   {journal} {Physica A: Statistical Mechanics and its Applications}\ }\textbf
  {\bibinfo {volume} {600}},\ \bibinfo {pages} {127538} (\bibinfo {year}
  {2022})}\BibitemShut {NoStop}%
\bibitem [{\citenamefont {Carrasquilla}\ and\ \citenamefont
  {Melko}(2017)}]{Carrasquilla_Melko_2017}%
  \BibitemOpen
  \bibfield  {author} {\bibinfo {author} {\bibfnamefont {J.}~\bibnamefont
  {Carrasquilla}}\ and\ \bibinfo {author} {\bibfnamefont {R.~G.}\ \bibnamefont
  {Melko}},\ }\bibfield  {title} {\bibinfo {title} {Machine learning phases of
  matter},\ }\href {https://doi.org/10.1038/nphys4035} {\bibfield  {journal}
  {\bibinfo  {journal} {Nature Physics}\ }\textbf {\bibinfo {volume} {13}},\
  \bibinfo {pages} {431–434} (\bibinfo {year} {2017})}\BibitemShut {NoStop}%
\bibitem [{\citenamefont {Deng}\ \emph {et~al.}(2017)\citenamefont {Deng},
  \citenamefont {Li},\ and\ \citenamefont
  {Das~Sarma}}]{DengLiDasSarma2017_MLTopStates}%
  \BibitemOpen
  \bibfield  {author} {\bibinfo {author} {\bibfnamefont {D.-L.}\ \bibnamefont
  {Deng}}, \bibinfo {author} {\bibfnamefont {X.}~\bibnamefont {Li}},\ and\
  \bibinfo {author} {\bibfnamefont {S.}~\bibnamefont {Das~Sarma}},\ }\bibfield
  {title} {\bibinfo {title} {Machine learning topological states},\ }\href
  {https://doi.org/10.1103/PhysRevB.96.195145} {\bibfield  {journal} {\bibinfo
  {journal} {Physical Review B}\ }\textbf {\bibinfo {volume} {96}},\ \bibinfo
  {pages} {195145} (\bibinfo {year} {2017})}\BibitemShut {NoStop}%
\bibitem [{\citenamefont {Zhang}\ \emph {et~al.}(2018)\citenamefont {Zhang},
  \citenamefont {Shen},\ and\ \citenamefont
  {Zhai}}]{ZhangShenZhai2018_TopInvNN}%
  \BibitemOpen
  \bibfield  {author} {\bibinfo {author} {\bibfnamefont {P.}~\bibnamefont
  {Zhang}}, \bibinfo {author} {\bibfnamefont {H.}~\bibnamefont {Shen}},\ and\
  \bibinfo {author} {\bibfnamefont {H.}~\bibnamefont {Zhai}},\ }\bibfield
  {title} {\bibinfo {title} {Machine learning topological invariants with
  neural networks},\ }\href {https://doi.org/10.1103/PhysRevLett.120.066401}
  {\bibfield  {journal} {\bibinfo  {journal} {Physical Review Letters}\
  }\textbf {\bibinfo {volume} {120}},\ \bibinfo {pages} {066401} (\bibinfo
  {year} {2018})}\BibitemShut {NoStop}%
\bibitem [{\citenamefont {van Nieuwenburg}\ \emph {et~al.}(2017)\citenamefont
  {van Nieuwenburg}, \citenamefont {Liu},\ and\ \citenamefont
  {Huber}}]{vNLH-confusion}%
  \BibitemOpen
  \bibfield  {author} {\bibinfo {author} {\bibfnamefont {E.~P.~L.}\
  \bibnamefont {van Nieuwenburg}}, \bibinfo {author} {\bibfnamefont {Y.-H.}\
  \bibnamefont {Liu}},\ and\ \bibinfo {author} {\bibfnamefont {S.~D.}\
  \bibnamefont {Huber}},\ }\bibfield  {title} {\bibinfo {title} {Learning phase
  transitions by confusion},\ }\href {https://doi.org/10.1038/nphys4037}
  {\bibfield  {journal} {\bibinfo  {journal} {Nature Physics}\ }\textbf
  {\bibinfo {volume} {13}},\ \bibinfo {pages} {435} (\bibinfo {year}
  {2017})}\BibitemShut {NoStop}%
\bibitem [{\citenamefont {Zhang}\ \emph {et~al.}(2020)\citenamefont {Zhang},
  \citenamefont {Ginsparg},\ and\ \citenamefont
  {Kim}}]{Zhang2020PhysRevResearch.2.023283}%
  \BibitemOpen
  \bibfield  {author} {\bibinfo {author} {\bibfnamefont {Y.}~\bibnamefont
  {Zhang}}, \bibinfo {author} {\bibfnamefont {P.}~\bibnamefont {Ginsparg}},\
  and\ \bibinfo {author} {\bibfnamefont {E.-A.}\ \bibnamefont {Kim}},\
  }\bibfield  {title} {\bibinfo {title} {Interpreting machine learning of
  topological quantum phase transitions},\ }\href
  {https://doi.org/10.1103/PhysRevResearch.2.023283} {\bibfield  {journal}
  {\bibinfo  {journal} {Phys. Rev. Res.}\ }\textbf {\bibinfo {volume} {2}},\
  \bibinfo {pages} {023283} (\bibinfo {year} {2020})}\BibitemShut {NoStop}%
\bibitem [{\citenamefont {Mondragon-Shem}\ \emph {et~al.}(2014)\citenamefont
  {Mondragon-Shem}, \citenamefont {Hughes}, \citenamefont {Song},\ and\
  \citenamefont {Prodan}}]{mondragon2014topological}%
  \BibitemOpen
  \bibfield  {author} {\bibinfo {author} {\bibfnamefont {I.}~\bibnamefont
  {Mondragon-Shem}}, \bibinfo {author} {\bibfnamefont {T.~L.}\ \bibnamefont
  {Hughes}}, \bibinfo {author} {\bibfnamefont {J.}~\bibnamefont {Song}},\ and\
  \bibinfo {author} {\bibfnamefont {E.}~\bibnamefont {Prodan}},\ }\bibfield
  {title} {\bibinfo {title} {Topological criticality in the chiral-symmetric
  aiii class at strong disorder},\ }\href
  {https://doi.org/10.1103/PhysRevLett.113.046802} {\bibfield  {journal}
  {\bibinfo  {journal} {Physical Review Letters}\ }\textbf {\bibinfo {volume}
  {113}},\ \bibinfo {pages} {046802} (\bibinfo {year} {2014})}\BibitemShut
  {NoStop}%
\bibitem [{\citenamefont {Chiu}\ \emph {et~al.}(2016)\citenamefont {Chiu},
  \citenamefont {Teo}, \citenamefont {Schnyder},\ and\ \citenamefont
  {Ryu}}]{Chiu2016RevModPhys.88.035005}%
  \BibitemOpen
  \bibfield  {author} {\bibinfo {author} {\bibfnamefont {C.-K.}\ \bibnamefont
  {Chiu}}, \bibinfo {author} {\bibfnamefont {J.~C.~Y.}\ \bibnamefont {Teo}},
  \bibinfo {author} {\bibfnamefont {A.~P.}\ \bibnamefont {Schnyder}},\ and\
  \bibinfo {author} {\bibfnamefont {S.}~\bibnamefont {Ryu}},\ }\bibfield
  {title} {\bibinfo {title} {Classification of topological quantum matter with
  symmetries},\ }\href {https://doi.org/10.1103/RevModPhys.88.035005}
  {\bibfield  {journal} {\bibinfo  {journal} {Rev. Mod. Phys.}\ }\textbf
  {\bibinfo {volume} {88}},\ \bibinfo {pages} {035005} (\bibinfo {year}
  {2016})}\BibitemShut {NoStop}%
\bibitem [{\citenamefont {Hasan}\ and\ \citenamefont
  {Kane}(2010)}]{HasanKane2010}%
  \BibitemOpen
  \bibfield  {author} {\bibinfo {author} {\bibfnamefont {M.~Z.}\ \bibnamefont
  {Hasan}}\ and\ \bibinfo {author} {\bibfnamefont {C.~L.}\ \bibnamefont
  {Kane}},\ }\bibfield  {title} {\bibinfo {title} {Colloquium: Topological
  insulators},\ }\href {https://doi.org/10.1103/RevModPhys.82.3045} {\bibfield
  {journal} {\bibinfo  {journal} {Reviews of Modern Physics}\ }\textbf
  {\bibinfo {volume} {82}},\ \bibinfo {pages} {3045} (\bibinfo {year}
  {2010})}\BibitemShut {NoStop}%
\bibitem [{\citenamefont {Qi}\ and\ \citenamefont {Zhang}(2011)}]{QiZhang2011}%
  \BibitemOpen
  \bibfield  {author} {\bibinfo {author} {\bibfnamefont {X.-L.}\ \bibnamefont
  {Qi}}\ and\ \bibinfo {author} {\bibfnamefont {S.-C.}\ \bibnamefont {Zhang}},\
  }\bibfield  {title} {\bibinfo {title} {Topological insulators and
  superconductors},\ }\href {https://doi.org/10.1103/RevModPhys.83.1057}
  {\bibfield  {journal} {\bibinfo  {journal} {Reviews of Modern Physics}\
  }\textbf {\bibinfo {volume} {83}},\ \bibinfo {pages} {1057} (\bibinfo {year}
  {2011})}\BibitemShut {NoStop}%
\bibitem [{\citenamefont {Bernevig}\ \emph {et~al.}(2006)\citenamefont
  {Bernevig}, \citenamefont {Hughes},\ and\ \citenamefont
  {Zhang}}]{BernevigHughesZhang2006}%
  \BibitemOpen
  \bibfield  {author} {\bibinfo {author} {\bibfnamefont {B.~A.}\ \bibnamefont
  {Bernevig}}, \bibinfo {author} {\bibfnamefont {T.~L.}\ \bibnamefont
  {Hughes}},\ and\ \bibinfo {author} {\bibfnamefont {S.-C.}\ \bibnamefont
  {Zhang}},\ }\bibfield  {title} {\bibinfo {title} {Quantum spin {H}all effect
  and topological phase transition in {H}g{T}e quantum wells},\ }\href
  {https://doi.org/10.1126/science.1133734} {\bibfield  {journal} {\bibinfo
  {journal} {Science}\ }\textbf {\bibinfo {volume} {314}},\ \bibinfo {pages}
  {1757} (\bibinfo {year} {2006})}\BibitemShut {NoStop}%
\bibitem [{\citenamefont {Fu}(2011)}]{Fu2011}%
  \BibitemOpen
  \bibfield  {author} {\bibinfo {author} {\bibfnamefont {L.}~\bibnamefont
  {Fu}},\ }\bibfield  {title} {\bibinfo {title} {Topological crystalline
  insulators},\ }\href {https://doi.org/10.1103/PhysRevLett.106.106802}
  {\bibfield  {journal} {\bibinfo  {journal} {Physical Review Letters}\
  }\textbf {\bibinfo {volume} {106}},\ \bibinfo {pages} {106802} (\bibinfo
  {year} {2011})}\BibitemShut {NoStop}%
\bibitem [{\citenamefont {Zhang}\ \emph {et~al.}(2009)\citenamefont {Zhang},
  \citenamefont {Liu}, \citenamefont {Qi}, \citenamefont {Dai}, \citenamefont
  {Fang},\ and\ \citenamefont {Zhang}}]{Zhang2009}%
  \BibitemOpen
  \bibfield  {author} {\bibinfo {author} {\bibfnamefont {H.~J.}\ \bibnamefont
  {Zhang}}, \bibinfo {author} {\bibfnamefont {C.~X.}\ \bibnamefont {Liu}},
  \bibinfo {author} {\bibfnamefont {X.~L.}\ \bibnamefont {Qi}}, \bibinfo
  {author} {\bibfnamefont {X.}~\bibnamefont {Dai}}, \bibinfo {author}
  {\bibfnamefont {Z.}~\bibnamefont {Fang}},\ and\ \bibinfo {author}
  {\bibfnamefont {S.-C.}\ \bibnamefont {Zhang}},\ }\bibfield  {title} {\bibinfo
  {title} {Topological insulators in {B}i$_{2}${S}e$_{3}$, {B}i$_{2}${T}e$_{3}$
  and {S}b$_{2}${T}e$_{3}$ with a single {D}irac cone on the surface},\ }\href
  {https://doi.org/10.1038/nphys1270} {\bibfield  {journal} {\bibinfo
  {journal} {Nature Physics}\ }\textbf {\bibinfo {volume} {5}},\ \bibinfo
  {pages} {438} (\bibinfo {year} {2009})}\BibitemShut {NoStop}%
\bibitem [{\citenamefont {Yan}\ and\ \citenamefont
  {Zhang}(2012)}]{YanZhang2012}%
  \BibitemOpen
  \bibfield  {author} {\bibinfo {author} {\bibfnamefont {B.}~\bibnamefont
  {Yan}}\ and\ \bibinfo {author} {\bibfnamefont {S.-C.}\ \bibnamefont
  {Zhang}},\ }\bibfield  {title} {\bibinfo {title} {Topological materials},\
  }\href {https://doi.org/10.1088/0034-4885/75/9/096501} {\bibfield  {journal}
  {\bibinfo  {journal} {Reports on Progress in Physics}\ }\textbf {\bibinfo
  {volume} {75}},\ \bibinfo {pages} {096501} (\bibinfo {year}
  {2012})}\BibitemShut {NoStop}%
\bibitem [{\citenamefont {Rachel}(2018)}]{Rachel2018}%
  \BibitemOpen
  \bibfield  {author} {\bibinfo {author} {\bibfnamefont {S.}~\bibnamefont
  {Rachel}},\ }\bibfield  {title} {\bibinfo {title} {Interacting topological
  insulators: a review},\ }\href {https://doi.org/10.1088/1361-6633/aad6a6}
  {\bibfield  {journal} {\bibinfo  {journal} {Reports on Progress in Physics}\
  }\textbf {\bibinfo {volume} {81}},\ \bibinfo {pages} {116501} (\bibinfo
  {year} {2018})}\BibitemShut {NoStop}%
\bibitem [{\citenamefont {Ando}\ and\ \citenamefont {Fu}(2015)}]{AndoFu2015}%
  \BibitemOpen
  \bibfield  {author} {\bibinfo {author} {\bibfnamefont {Y.}~\bibnamefont
  {Ando}}\ and\ \bibinfo {author} {\bibfnamefont {L.}~\bibnamefont {Fu}},\
  }\bibfield  {title} {\bibinfo {title} {Topological crystalline insulators and
  topological superconductors: From concepts to materials},\ }\href
  {https://doi.org/10.1146/annurev-conmatphys-031214-014501} {\bibfield
  {journal} {\bibinfo  {journal} {Annual Review of Condensed Matter Physics}\
  }\textbf {\bibinfo {volume} {6}},\ \bibinfo {pages} {361} (\bibinfo {year}
  {2015})}\BibitemShut {NoStop}%
\bibitem [{\citenamefont {De~L{\'e}s{\'e}leuc}\ \emph
  {et~al.}(2019)\citenamefont {De~L{\'e}s{\'e}leuc}, \citenamefont {Lienhard},
  \citenamefont {Scholl}, \citenamefont {Barredo}, \citenamefont {Weber},
  \citenamefont {Lang}, \citenamefont {B{\"u}chler}, \citenamefont {Lahaye},\
  and\ \citenamefont {Browaeys}}]{de2019observation}%
  \BibitemOpen
  \bibfield  {author} {\bibinfo {author} {\bibfnamefont {S.}~\bibnamefont
  {De~L{\'e}s{\'e}leuc}}, \bibinfo {author} {\bibfnamefont {V.}~\bibnamefont
  {Lienhard}}, \bibinfo {author} {\bibfnamefont {P.}~\bibnamefont {Scholl}},
  \bibinfo {author} {\bibfnamefont {D.}~\bibnamefont {Barredo}}, \bibinfo
  {author} {\bibfnamefont {S.}~\bibnamefont {Weber}}, \bibinfo {author}
  {\bibfnamefont {N.}~\bibnamefont {Lang}}, \bibinfo {author} {\bibfnamefont
  {H.~P.}\ \bibnamefont {B{\"u}chler}}, \bibinfo {author} {\bibfnamefont
  {T.}~\bibnamefont {Lahaye}},\ and\ \bibinfo {author} {\bibfnamefont
  {A.}~\bibnamefont {Browaeys}},\ }\bibfield  {title} {\bibinfo {title}
  {Observation of a symmetry-protected topological phase of interacting bosons
  with {R}ydberg atoms},\ }\href {https://doi.org/10.1126/science.aav9105}
  {\bibfield  {journal} {\bibinfo  {journal} {Science}\ }\textbf {\bibinfo
  {volume} {365}},\ \bibinfo {pages} {775} (\bibinfo {year}
  {2019})}\BibitemShut {NoStop}%
\bibitem [{\citenamefont {Splitthoff}\ \emph {et~al.}(2024)\citenamefont
  {Splitthoff}, \citenamefont {Belo}, \citenamefont {Jin}, \citenamefont {Liu},
  \citenamefont {Greplova},\ and\ \citenamefont
  {Andersen}}]{splitthoff2024gate}%
  \BibitemOpen
  \bibfield  {author} {\bibinfo {author} {\bibfnamefont {L.~J.}\ \bibnamefont
  {Splitthoff}}, \bibinfo {author} {\bibfnamefont {M.~C.}\ \bibnamefont
  {Belo}}, \bibinfo {author} {\bibfnamefont {G.}~\bibnamefont {Jin}}, \bibinfo
  {author} {\bibfnamefont {Y.}~\bibnamefont {Liu}}, \bibinfo {author}
  {\bibfnamefont {E.}~\bibnamefont {Greplova}},\ and\ \bibinfo {author}
  {\bibfnamefont {C.~K.}\ \bibnamefont {Andersen}},\ }\bibfield  {title}
  {\bibinfo {title} {Gate-tunable phase transition in a resonator-based
  {S}u-{S}chrieffer-{H}eeger chain},\ }\href
  {https://doi.org/10.1103/PhysRevResearch.6.043286} {\bibfield  {journal}
  {\bibinfo  {journal} {Physical Review Research}\ }\textbf {\bibinfo {volume}
  {6}},\ \bibinfo {pages} {043286} (\bibinfo {year} {2024})}\BibitemShut
  {NoStop}%
\bibitem [{\citenamefont {Kanungo}\ \emph {et~al.}(2022)\citenamefont
  {Kanungo}, \citenamefont {Whalen}, \citenamefont {Lu}, \citenamefont {Yuan},
  \citenamefont {Dasgupta}, \citenamefont {Dunning}, \citenamefont {Hazzard},\
  and\ \citenamefont {Killian}}]{Kanungo_2022}%
  \BibitemOpen
  \bibfield  {author} {\bibinfo {author} {\bibfnamefont {S.~K.}\ \bibnamefont
  {Kanungo}}, \bibinfo {author} {\bibfnamefont {J.~D.}\ \bibnamefont {Whalen}},
  \bibinfo {author} {\bibfnamefont {Y.}~\bibnamefont {Lu}}, \bibinfo {author}
  {\bibfnamefont {M.}~\bibnamefont {Yuan}}, \bibinfo {author} {\bibfnamefont
  {S.}~\bibnamefont {Dasgupta}}, \bibinfo {author} {\bibfnamefont {F.~B.}\
  \bibnamefont {Dunning}}, \bibinfo {author} {\bibfnamefont {K.~R.~A.}\
  \bibnamefont {Hazzard}},\ and\ \bibinfo {author} {\bibfnamefont {T.~C.}\
  \bibnamefont {Killian}},\ }\bibfield  {title} {\bibinfo {title} {Realizing
  topological edge states with {R}ydberg-atom synthetic dimensions},\ }\href
  {https://doi.org/10.1038/s41467-022-28550-y} {\bibfield  {journal} {\bibinfo
  {journal} {Nature Communications}\ }\textbf {\bibinfo {volume} {13}},\
  \bibinfo {pages} {972} (\bibinfo {year} {2022})}\BibitemShut {NoStop}%
\bibitem [{\citenamefont {Kiczynski}\ \emph {et~al.}(2022)\citenamefont
  {Kiczynski}, \citenamefont {Gorman}, \citenamefont {Geng}, \citenamefont
  {Donnelly}, \citenamefont {Chung}, \citenamefont {He}, \citenamefont
  {Keizer},\ and\ \citenamefont {Simmons}}]{kiczynski2022engineering}%
  \BibitemOpen
  \bibfield  {author} {\bibinfo {author} {\bibfnamefont {M.}~\bibnamefont
  {Kiczynski}}, \bibinfo {author} {\bibfnamefont {S.}~\bibnamefont {Gorman}},
  \bibinfo {author} {\bibfnamefont {H.}~\bibnamefont {Geng}}, \bibinfo {author}
  {\bibfnamefont {M.}~\bibnamefont {Donnelly}}, \bibinfo {author}
  {\bibfnamefont {Y.}~\bibnamefont {Chung}}, \bibinfo {author} {\bibfnamefont
  {Y.}~\bibnamefont {He}}, \bibinfo {author} {\bibfnamefont {J.}~\bibnamefont
  {Keizer}},\ and\ \bibinfo {author} {\bibfnamefont {M.}~\bibnamefont
  {Simmons}},\ }\bibfield  {title} {\bibinfo {title} {Engineering topological
  states in atom-based semiconductor quantum dots},\ }\href
  {https://doi.org/10.1038/s41586-022-04706-0} {\bibfield  {journal} {\bibinfo
  {journal} {Nature}\ }\textbf {\bibinfo {volume} {606}},\ \bibinfo {pages}
  {694} (\bibinfo {year} {2022})}\BibitemShut {NoStop}%
\bibitem [{\citenamefont {Jouanny}\ \emph {et~al.}(2025)\citenamefont
  {Jouanny}, \citenamefont {Frasca}, \citenamefont {Weibel}, \citenamefont
  {Peyruchat}, \citenamefont {Scigliuzzo}, \citenamefont {Oppliger},
  \citenamefont {De~Palma}, \citenamefont {Sbroggio}, \citenamefont {Beaulieu},
  \citenamefont {Zilberberg} \emph {et~al.}}]{jouanny2024band}%
  \BibitemOpen
  \bibfield  {author} {\bibinfo {author} {\bibfnamefont {V.}~\bibnamefont
  {Jouanny}}, \bibinfo {author} {\bibfnamefont {S.}~\bibnamefont {Frasca}},
  \bibinfo {author} {\bibfnamefont {V.~J.}\ \bibnamefont {Weibel}}, \bibinfo
  {author} {\bibfnamefont {L.}~\bibnamefont {Peyruchat}}, \bibinfo {author}
  {\bibfnamefont {M.}~\bibnamefont {Scigliuzzo}}, \bibinfo {author}
  {\bibfnamefont {F.}~\bibnamefont {Oppliger}}, \bibinfo {author}
  {\bibfnamefont {F.}~\bibnamefont {De~Palma}}, \bibinfo {author}
  {\bibfnamefont {D.}~\bibnamefont {Sbroggio}}, \bibinfo {author}
  {\bibfnamefont {G.}~\bibnamefont {Beaulieu}}, \bibinfo {author}
  {\bibfnamefont {O.}~\bibnamefont {Zilberberg}}, \emph {et~al.},\ }\bibfield
  {title} {\bibinfo {title} {Band engineering and study of disorder using
  topology in compact high kinetic inductance cavity arrays},\ }\href
  {https://doi.org/10.1038/s41467-025-58595-8} {\bibfield  {journal} {\bibinfo
  {journal} {Nature Communications}\ }\textbf {\bibinfo {volume} {16}},\
  \bibinfo {pages} {3396} (\bibinfo {year} {2025})}\BibitemShut {NoStop}%
\bibitem [{\citenamefont {Mei}\ \emph {et~al.}(2018)\citenamefont {Mei},
  \citenamefont {Chen}, \citenamefont {Tian}, \citenamefont {Zhu},\ and\
  \citenamefont {Jia}}]{mei2018robust}%
  \BibitemOpen
  \bibfield  {author} {\bibinfo {author} {\bibfnamefont {F.}~\bibnamefont
  {Mei}}, \bibinfo {author} {\bibfnamefont {G.}~\bibnamefont {Chen}}, \bibinfo
  {author} {\bibfnamefont {L.}~\bibnamefont {Tian}}, \bibinfo {author}
  {\bibfnamefont {S.-L.}\ \bibnamefont {Zhu}},\ and\ \bibinfo {author}
  {\bibfnamefont {S.}~\bibnamefont {Jia}},\ }\bibfield  {title} {\bibinfo
  {title} {Robust quantum state transfer via topological edge states in
  superconducting qubit chains},\ }\href
  {https://doi.org/10.1103/physreva.98.012331} {\bibfield  {journal} {\bibinfo
  {journal} {Physical Review A}\ }\textbf {\bibinfo {volume} {98}},\ \bibinfo
  {pages} {012331} (\bibinfo {year} {2018})}\BibitemShut {NoStop}%
\bibitem [{\citenamefont {Zheng}\ \emph {et~al.}(2022)\citenamefont {Zheng},
  \citenamefont {Yi},\ and\ \citenamefont {Wang}}]{zheng2022engineering}%
  \BibitemOpen
  \bibfield  {author} {\bibinfo {author} {\bibfnamefont {L.-N.}\ \bibnamefont
  {Zheng}}, \bibinfo {author} {\bibfnamefont {X.}~\bibnamefont {Yi}},\ and\
  \bibinfo {author} {\bibfnamefont {H.-F.}\ \bibnamefont {Wang}},\ }\bibfield
  {title} {\bibinfo {title} {Engineering a phase-robust topological router in a
  dimerized superconducting-circuit lattice with long-range hopping and chiral
  symmetry},\ }\href {https://doi.org/10.1103/physrevapplied.18.054037}
  {\bibfield  {journal} {\bibinfo  {journal} {Physical Review Applied}\
  }\textbf {\bibinfo {volume} {18}},\ \bibinfo {pages} {054037} (\bibinfo
  {year} {2022})}\BibitemShut {NoStop}%
\bibitem [{\citenamefont {Kim}\ \emph {et~al.}(2021)\citenamefont {Kim},
  \citenamefont {Zhang}, \citenamefont {Ferreira}, \citenamefont {Banker},
  \citenamefont {Iverson}, \citenamefont {Sipahigil}, \citenamefont {Bello},
  \citenamefont {Gonz{\'a}lez-Tudela}, \citenamefont {Mirhosseini},\ and\
  \citenamefont {Painter}}]{kim2021quantum}%
  \BibitemOpen
  \bibfield  {author} {\bibinfo {author} {\bibfnamefont {E.}~\bibnamefont
  {Kim}}, \bibinfo {author} {\bibfnamefont {X.}~\bibnamefont {Zhang}}, \bibinfo
  {author} {\bibfnamefont {V.~S.}\ \bibnamefont {Ferreira}}, \bibinfo {author}
  {\bibfnamefont {J.}~\bibnamefont {Banker}}, \bibinfo {author} {\bibfnamefont
  {J.~K.}\ \bibnamefont {Iverson}}, \bibinfo {author} {\bibfnamefont
  {A.}~\bibnamefont {Sipahigil}}, \bibinfo {author} {\bibfnamefont
  {M.}~\bibnamefont {Bello}}, \bibinfo {author} {\bibfnamefont
  {A.}~\bibnamefont {Gonz{\'a}lez-Tudela}}, \bibinfo {author} {\bibfnamefont
  {M.}~\bibnamefont {Mirhosseini}},\ and\ \bibinfo {author} {\bibfnamefont
  {O.}~\bibnamefont {Painter}},\ }\bibfield  {title} {\bibinfo {title} {Quantum
  electrodynamics in a topological waveguide},\ }\href
  {https://doi.org/10.1103/physrevx.11.011015} {\bibfield  {journal} {\bibinfo
  {journal} {Physical Review X}\ }\textbf {\bibinfo {volume} {11}},\ \bibinfo
  {pages} {011015} (\bibinfo {year} {2021})}\BibitemShut {NoStop}%
\bibitem [{\citenamefont {Vega}\ \emph {et~al.}(2021)\citenamefont {Vega},
  \citenamefont {Bello}, \citenamefont {Porras},\ and\ \citenamefont
  {Gonz{\'a}lez-Tudela}}]{vega2021qubit}%
  \BibitemOpen
  \bibfield  {author} {\bibinfo {author} {\bibfnamefont {C.}~\bibnamefont
  {Vega}}, \bibinfo {author} {\bibfnamefont {M.}~\bibnamefont {Bello}},
  \bibinfo {author} {\bibfnamefont {D.}~\bibnamefont {Porras}},\ and\ \bibinfo
  {author} {\bibfnamefont {A.}~\bibnamefont {Gonz{\'a}lez-Tudela}},\ }\bibfield
   {title} {\bibinfo {title} {Qubit-photon bound states in topological
  waveguides with long-range hoppings},\ }\href
  {https://doi.org/10.1103/physreva.104.053522} {\bibfield  {journal} {\bibinfo
   {journal} {Physical Review A}\ }\textbf {\bibinfo {volume} {104}},\ \bibinfo
  {pages} {053522} (\bibinfo {year} {2021})}\BibitemShut {NoStop}%
\bibitem [{\citenamefont {Altland}\ and\ \citenamefont
  {Zirnbauer}(1997)}]{AltlandZirnbauer1997_AZ}%
  \BibitemOpen
  \bibfield  {author} {\bibinfo {author} {\bibfnamefont {A.}~\bibnamefont
  {Altland}}\ and\ \bibinfo {author} {\bibfnamefont {M.~R.}\ \bibnamefont
  {Zirnbauer}},\ }\bibfield  {title} {\bibinfo {title} {Nonstandard symmetry
  classes in mesoscopic normal-superconducting hybrid structures},\ }\href
  {https://doi.org/10.1103/PhysRevB.55.1142} {\bibfield  {journal} {\bibinfo
  {journal} {Physical Review B}\ }\textbf {\bibinfo {volume} {55}},\ \bibinfo
  {pages} {1142} (\bibinfo {year} {1997})}\BibitemShut {NoStop}%
\bibitem [{\citenamefont {Schnyder}\ \emph {et~al.}(2008)\citenamefont
  {Schnyder}, \citenamefont {Ryu}, \citenamefont {Furusaki},\ and\
  \citenamefont {Ludwig}}]{SchnyderRyuFurusakiLudwig2008}%
  \BibitemOpen
  \bibfield  {author} {\bibinfo {author} {\bibfnamefont {A.~P.}\ \bibnamefont
  {Schnyder}}, \bibinfo {author} {\bibfnamefont {S.}~\bibnamefont {Ryu}},
  \bibinfo {author} {\bibfnamefont {A.}~\bibnamefont {Furusaki}},\ and\
  \bibinfo {author} {\bibfnamefont {A.~W.~W.}\ \bibnamefont {Ludwig}},\
  }\bibfield  {title} {\bibinfo {title} {Classification of topological
  insulators and superconductors in three spatial dimensions},\ }\href
  {https://doi.org/10.1103/PhysRevB.78.195125} {\bibfield  {journal} {\bibinfo
  {journal} {Physical Review B}\ }\textbf {\bibinfo {volume} {78}},\ \bibinfo
  {pages} {195125} (\bibinfo {year} {2008})}\BibitemShut {NoStop}%
\bibitem [{\citenamefont {Kitaev}(2009)}]{Kitaev2009_PeriodicTable}%
  \BibitemOpen
  \bibfield  {author} {\bibinfo {author} {\bibfnamefont {A.}~\bibnamefont
  {Kitaev}},\ }\bibfield  {title} {\bibinfo {title} {Periodic table for
  topological insulators and superconductors},\ }in\ \href
  {https://doi.org/10.1063/1.3149495} {\emph {\bibinfo {booktitle} {AIP
  Conference Proceedings}}},\ Vol.\ \bibinfo {volume} {1134}\ (\bibinfo {year}
  {2009})\ pp.\ \bibinfo {pages} {22--30}\BibitemShut {NoStop}%
\bibitem [{\citenamefont {Jin}\ \emph {et~al.}(2024)\citenamefont {Jin},
  \citenamefont {Oriekhov}, \citenamefont {Splitthoff},\ and\ \citenamefont
  {Greplova}}]{jin2024topologicalfinitesizeeffect}%
  \BibitemOpen
  \bibfield  {author} {\bibinfo {author} {\bibfnamefont {G.}~\bibnamefont
  {Jin}}, \bibinfo {author} {\bibfnamefont {D.~O.}\ \bibnamefont {Oriekhov}},
  \bibinfo {author} {\bibfnamefont {L.~J.}\ \bibnamefont {Splitthoff}},\ and\
  \bibinfo {author} {\bibfnamefont {E.}~\bibnamefont {Greplova}},\ }\bibfield
  {title} {\bibinfo {title} {Topological finite size effect in one-dimensional
  chiral symmetric systems},\ }\href
  {https://doi.org/10.48550/arXiv.2411.17822} {\bibfield  {journal} {\bibinfo
  {journal} {arxiv:2411.17822}\ } (\bibinfo {year} {2024})}\BibitemShut
  {NoStop}%
\bibitem [{\citenamefont {Greplova}\ \emph {et~al.}(2020)\citenamefont
  {Greplova}, \citenamefont {Valenti}, \citenamefont {Boschung}, \citenamefont
  {Sch\"{a}fer}, \citenamefont {L\"{o}rch},\ and\ \citenamefont
  {Huber}}]{Greplova_Valenti_Boschung_Huber_2020}%
  \BibitemOpen
  \bibfield  {author} {\bibinfo {author} {\bibfnamefont {E.}~\bibnamefont
  {Greplova}}, \bibinfo {author} {\bibfnamefont {A.}~\bibnamefont {Valenti}},
  \bibinfo {author} {\bibfnamefont {G.}~\bibnamefont {Boschung}}, \bibinfo
  {author} {\bibfnamefont {F.}~\bibnamefont {Sch\"{a}fer}}, \bibinfo {author}
  {\bibfnamefont {N.}~\bibnamefont {L\"{o}rch}},\ and\ \bibinfo {author}
  {\bibfnamefont {S.~D.}\ \bibnamefont {Huber}},\ }\bibfield  {title} {\bibinfo
  {title} {Unsupervised identification of topological phase transitions using
  predictive models},\ }\href {https://doi.org/10.1088/1367-2630/ab7771}
  {\bibfield  {journal} {\bibinfo  {journal} {New Journal of Physics}\ }\textbf
  {\bibinfo {volume} {22}},\ \bibinfo {pages} {045003} (\bibinfo {year}
  {2020})}\BibitemShut {NoStop}%
\bibitem [{\citenamefont {Sun}\ \emph {et~al.}(2018)\citenamefont {Sun},
  \citenamefont {Yi}, \citenamefont {Zhang}, \citenamefont {Shen},\ and\
  \citenamefont {Zhai}}]{Sun2018PhysRevB.98.085402}%
  \BibitemOpen
  \bibfield  {author} {\bibinfo {author} {\bibfnamefont {N.}~\bibnamefont
  {Sun}}, \bibinfo {author} {\bibfnamefont {J.}~\bibnamefont {Yi}}, \bibinfo
  {author} {\bibfnamefont {P.}~\bibnamefont {Zhang}}, \bibinfo {author}
  {\bibfnamefont {H.}~\bibnamefont {Shen}},\ and\ \bibinfo {author}
  {\bibfnamefont {H.}~\bibnamefont {Zhai}},\ }\bibfield  {title} {\bibinfo
  {title} {Deep learning topological invariants of band insulators},\ }\href
  {https://doi.org/10.1103/PhysRevB.98.085402} {\bibfield  {journal} {\bibinfo
  {journal} {Physical Review B}\ }\textbf {\bibinfo {volume} {98}},\ \bibinfo
  {pages} {085402} (\bibinfo {year} {2018})}\BibitemShut {NoStop}%
\bibitem [{\citenamefont {Zhang}\ and\ \citenamefont
  {Kim}(2017)}]{Zhang2017PhysRevLett.118.216401}%
  \BibitemOpen
  \bibfield  {author} {\bibinfo {author} {\bibfnamefont {Y.}~\bibnamefont
  {Zhang}}\ and\ \bibinfo {author} {\bibfnamefont {E.-A.}\ \bibnamefont
  {Kim}},\ }\bibfield  {title} {\bibinfo {title} {Quantum loop topography for
  machine learning},\ }\href {https://doi.org/10.1103/PhysRevLett.118.216401}
  {\bibfield  {journal} {\bibinfo  {journal} {Physical Review Letters}\
  }\textbf {\bibinfo {volume} {118}},\ \bibinfo {pages} {216401} (\bibinfo
  {year} {2017})}\BibitemShut {NoStop}%
\bibitem [{\citenamefont {Baireuther}\ \emph {et~al.}(2023)\citenamefont
  {Baireuther}, \citenamefont {Płodzień}, \citenamefont {Ojanen},
  \citenamefont {Tworzydło},\ and\ \citenamefont {Hyart}}]{Baireuther2023}%
  \BibitemOpen
  \bibfield  {author} {\bibinfo {author} {\bibfnamefont {P.}~\bibnamefont
  {Baireuther}}, \bibinfo {author} {\bibfnamefont {M.}~\bibnamefont
  {Płodzień}}, \bibinfo {author} {\bibfnamefont {T.}~\bibnamefont {Ojanen}},
  \bibinfo {author} {\bibfnamefont {J.}~\bibnamefont {Tworzydło}},\ and\
  \bibinfo {author} {\bibfnamefont {T.}~\bibnamefont {Hyart}},\ }\bibfield
  {title} {\bibinfo {title} {Identifying chern numbers of superconductors from
  local measurements},\ }\href
  {https://doi.org/10.21468/scipostphyscore.6.4.087} {\bibfield  {journal}
  {\bibinfo  {journal} {SciPost Physics Core}\ }\textbf {\bibinfo {volume}
  {6}},\ \bibinfo {pages} {087} (\bibinfo {year} {2023})}\BibitemShut {NoStop}%
\bibitem [{\citenamefont {Caio}\ \emph {et~al.}(2019)\citenamefont {Caio},
  \citenamefont {Caccin}, \citenamefont {Baireuther}, \citenamefont {Hyart},\
  and\ \citenamefont {Fruchart}}]{caio2019machine}%
  \BibitemOpen
  \bibfield  {author} {\bibinfo {author} {\bibfnamefont {M.~D.}\ \bibnamefont
  {Caio}}, \bibinfo {author} {\bibfnamefont {M.}~\bibnamefont {Caccin}},
  \bibinfo {author} {\bibfnamefont {P.}~\bibnamefont {Baireuther}}, \bibinfo
  {author} {\bibfnamefont {T.}~\bibnamefont {Hyart}},\ and\ \bibinfo {author}
  {\bibfnamefont {M.}~\bibnamefont {Fruchart}},\ }\bibfield  {title} {\bibinfo
  {title} {Machine learning assisted measurement of local topological
  invariants},\ }\href
  {https://doi.org/https://doi.org/10.48550/arXiv.1901.03346} {\bibfield
  {journal} {\bibinfo  {journal} {arXiv:1901.03346}\ } (\bibinfo {year}
  {2019})}\BibitemShut {NoStop}%
\bibitem [{\citenamefont {Zhang}\ \emph {et~al.}(2021)\citenamefont {Zhang},
  \citenamefont {Tang}, \citenamefont {Huang}, \citenamefont {Zhang},
  \citenamefont {Huang},\ and\ \citenamefont {Zhang}}]{Zhang2021nonHerm}%
  \BibitemOpen
  \bibfield  {author} {\bibinfo {author} {\bibfnamefont {L.-F.}\ \bibnamefont
  {Zhang}}, \bibinfo {author} {\bibfnamefont {L.-Z.}\ \bibnamefont {Tang}},
  \bibinfo {author} {\bibfnamefont {Z.-H.}\ \bibnamefont {Huang}}, \bibinfo
  {author} {\bibfnamefont {G.-Q.}\ \bibnamefont {Zhang}}, \bibinfo {author}
  {\bibfnamefont {W.}~\bibnamefont {Huang}},\ and\ \bibinfo {author}
  {\bibfnamefont {D.-W.}\ \bibnamefont {Zhang}},\ }\bibfield  {title} {\bibinfo
  {title} {Machine learning topological invariants of non-{H}ermitian
  systems},\ }\href {https://doi.org/10.1103/physreva.103.012419} {\bibfield
  {journal} {\bibinfo  {journal} {Physical Review A}\ }\textbf {\bibinfo
  {volume} {103}},\ \bibinfo {pages} {012419} (\bibinfo {year}
  {2021})}\BibitemShut {NoStop}%
\bibitem [{\citenamefont {Molignini}\ \emph {et~al.}(2021)\citenamefont
  {Molignini}, \citenamefont {Zegarra}, \citenamefont {van Nieuwenburg},
  \citenamefont {Chitra},\ and\ \citenamefont
  {Chen}}]{Molignini2021PostPhys.11.3.073}%
  \BibitemOpen
  \bibfield  {author} {\bibinfo {author} {\bibfnamefont {P.}~\bibnamefont
  {Molignini}}, \bibinfo {author} {\bibfnamefont {A.}~\bibnamefont {Zegarra}},
  \bibinfo {author} {\bibfnamefont {E.}~\bibnamefont {van Nieuwenburg}},
  \bibinfo {author} {\bibfnamefont {R.}~\bibnamefont {Chitra}},\ and\ \bibinfo
  {author} {\bibfnamefont {W.}~\bibnamefont {Chen}},\ }\bibfield  {title}
  {\bibinfo {title} {{A supervised learning algorithm for interacting
  topological insulators based on local curvature}},\ }\href
  {https://doi.org/10.21468/SciPostPhys.11.3.073} {\bibfield  {journal}
  {\bibinfo  {journal} {SciPost Phys.}\ }\textbf {\bibinfo {volume} {11}},\
  \bibinfo {pages} {073} (\bibinfo {year} {2021})}\BibitemShut {NoStop}%
\bibitem [{\citenamefont {Ghosh}\ and\ \citenamefont
  {Sarkar}(2024)}]{Ghosh2024PRB}%
  \BibitemOpen
  \bibfield  {author} {\bibinfo {author} {\bibfnamefont {A.}~\bibnamefont
  {Ghosh}}\ and\ \bibinfo {author} {\bibfnamefont {M.}~\bibnamefont {Sarkar}},\
  }\bibfield  {title} {\bibinfo {title} {Supervised learning of an interacting
  two-dimensional hardcore boson model of a weak topological insulator using
  correlation functions},\ }\href {https://doi.org/10.1103/physrevb.110.165134}
  {\bibfield  {journal} {\bibinfo  {journal} {Physical Review B}\ }\textbf
  {\bibinfo {volume} {110}},\ \bibinfo {pages} {165134} (\bibinfo {year}
  {2024})}\BibitemShut {NoStop}%
\bibitem [{\citenamefont {Ghosh}\ \emph {et~al.}(2025)\citenamefont {Ghosh},
  \citenamefont {Sarkar}, \citenamefont {Kao},\ and\ \citenamefont
  {Chen}}]{Ghosh2025IOP}%
  \BibitemOpen
  \bibfield  {author} {\bibinfo {author} {\bibfnamefont {A.}~\bibnamefont
  {Ghosh}}, \bibinfo {author} {\bibfnamefont {M.}~\bibnamefont {Sarkar}},
  \bibinfo {author} {\bibfnamefont {Y.-J.}\ \bibnamefont {Kao}},\ and\ \bibinfo
  {author} {\bibfnamefont {P.}~\bibnamefont {Chen}},\ }\bibfield  {title}
  {\bibinfo {title} {Learning phases with quantum {M}onte {C}arlo simulation
  cell},\ }\href {https://doi.org/10.1088/2632-2153/ae107c} {\bibfield
  {journal} {\bibinfo  {journal} {Machine Learning: Science and Technology}\
  }\textbf {\bibinfo {volume} {6}},\ \bibinfo {pages} {045017} (\bibinfo {year}
  {2025})}\BibitemShut {NoStop}%
\bibitem [{Sup({\natexlab{a}})}]{Supplement}%
  \BibitemOpen
  \href@noop {} {} ({\natexlab{a}}),\ \bibinfo {note} {see Supplemental
  Material at [] for the derivation of of reduced real-space winding number
  formula, correlations analysis of SSH dataset \cite{StanThesis}, SVD pruning
  of the usual classifier \cite{BadrThesis}, classification task benchmarking,
  and description of used neural network architectures.}\BibitemShut {Stop}%
\bibitem [{\citenamefont {Su}\ \emph {et~al.}(1979)\citenamefont {Su},
  \citenamefont {Schrieffer},\ and\ \citenamefont {Heeger}}]{su1979solitons}%
  \BibitemOpen
  \bibfield  {author} {\bibinfo {author} {\bibfnamefont {W.-P.}\ \bibnamefont
  {Su}}, \bibinfo {author} {\bibfnamefont {J.~R.}\ \bibnamefont {Schrieffer}},\
  and\ \bibinfo {author} {\bibfnamefont {A.~J.}\ \bibnamefont {Heeger}},\
  }\bibfield  {title} {\bibinfo {title} {Solitons in polyacetylene},\ }\href
  {https://doi.org/10.1103/physrevlett.42.1698} {\bibfield  {journal} {\bibinfo
   {journal} {Physical Review Letters}\ }\textbf {\bibinfo {volume} {42}},\
  \bibinfo {pages} {1698} (\bibinfo {year} {1979})}\BibitemShut {NoStop}%
\bibitem [{\citenamefont {Asb{\'o}th}\ \emph {et~al.}(2016)\citenamefont
  {Asb{\'o}th}, \citenamefont {Oroszl{\'a}ny},\ and\ \citenamefont
  {P{\'a}lyi}}]{asboth2016short}%
  \BibitemOpen
  \bibfield  {author} {\bibinfo {author} {\bibfnamefont {J.~K.}\ \bibnamefont
  {Asb{\'o}th}}, \bibinfo {author} {\bibfnamefont {L.}~\bibnamefont
  {Oroszl{\'a}ny}},\ and\ \bibinfo {author} {\bibfnamefont {A.}~\bibnamefont
  {P{\'a}lyi}},\ }\bibfield  {title} {\bibinfo {title} {A short course on
  topological insulators},\ }\href {https://doi.org/10.1007/978-3-319-25607-8}
  {\bibfield  {journal} {\bibinfo  {journal} {Lecture notes in physics}\
  }\textbf {\bibinfo {volume} {919}} (\bibinfo {year} {2016})}\BibitemShut
  {NoStop}%
\bibitem [{\citenamefont {P{\'e}rez-Gonz{\'a}lez}\ \emph
  {et~al.}(2018)\citenamefont {P{\'e}rez-Gonz{\'a}lez}, \citenamefont {Bello},
  \citenamefont {G{\'o}mez-Le{\'o}n},\ and\ \citenamefont
  {Platero}}]{perez2018ssh}%
  \BibitemOpen
  \bibfield  {author} {\bibinfo {author} {\bibfnamefont {B.}~\bibnamefont
  {P{\'e}rez-Gonz{\'a}lez}}, \bibinfo {author} {\bibfnamefont {M.}~\bibnamefont
  {Bello}}, \bibinfo {author} {\bibfnamefont {{\'A}.}~\bibnamefont
  {G{\'o}mez-Le{\'o}n}},\ and\ \bibinfo {author} {\bibfnamefont
  {G.}~\bibnamefont {Platero}},\ }\bibfield  {title} {\bibinfo {title} {{SSH}
  model with long-range hoppings: topology, driving and disorder},\ }\href
  {https://doi.org/10.48550/arXiv.1802.03973} {\bibfield  {journal} {\bibinfo
  {journal} {arxiv:1802.03973}\ } (\bibinfo {year} {2018})}\BibitemShut
  {NoStop}%
\bibitem [{\citenamefont {Cybinski}\ \emph
  {et~al.}(2024{\natexlab{a}})\citenamefont {Cybinski}, \citenamefont
  {Plodzien}, \citenamefont {Tomza}, \citenamefont {Lewenstein}, \citenamefont
  {Dauphin},\ and\ \citenamefont {Dawid}}]{Cybinski2024}%
  \BibitemOpen
  \bibfield  {author} {\bibinfo {author} {\bibfnamefont {K.}~\bibnamefont
  {Cybinski}}, \bibinfo {author} {\bibfnamefont {M.}~\bibnamefont {Plodzien}},
  \bibinfo {author} {\bibfnamefont {M.}~\bibnamefont {Tomza}}, \bibinfo
  {author} {\bibfnamefont {M.}~\bibnamefont {Lewenstein}}, \bibinfo {author}
  {\bibfnamefont {A.}~\bibnamefont {Dauphin}},\ and\ \bibinfo {author}
  {\bibfnamefont {A.}~\bibnamefont {Dawid}},\ }\bibfield  {title} {\bibinfo
  {title} {Characterizing out-of-distribution generalization of neural
  networks: application to the disordered {S}u-{S}chrieffer-{H}eeger model},\
  }\href {https://doi.org/10.1088/2632-2153/ad9079} {\bibfield  {journal}
  {\bibinfo  {journal} {Machine Learning: Science and Technology}\ }\textbf
  {\bibinfo {volume} {6}},\ \bibinfo {pages} {015014} (\bibinfo {year}
  {2024}{\natexlab{a}})}\BibitemShut {NoStop}%
\bibitem [{\citenamefont {Tsang}\ \emph {et~al.}(2017)\citenamefont {Tsang},
  \citenamefont {Cheng},\ and\ \citenamefont
  {Liu}}]{Tsang2017_InteractionsWeights}%
  \BibitemOpen
  \bibfield  {author} {\bibinfo {author} {\bibfnamefont {M.}~\bibnamefont
  {Tsang}}, \bibinfo {author} {\bibfnamefont {D.}~\bibnamefont {Cheng}},\ and\
  \bibinfo {author} {\bibfnamefont {Y.}~\bibnamefont {Liu}},\ }\bibfield
  {title} {\bibinfo {title} {Detecting statistical interactions from neural
  network weights},\ }\href {https://doi.org/10.48550/arXiv.1705.04977}
  {\bibfield  {journal} {\bibinfo  {journal} {arxiv:1705.04977}\ } (\bibinfo
  {year} {2017})}\BibitemShut {NoStop}%
\bibitem [{\citenamefont {Eilertsen}\ \emph {et~al.}(2020)\citenamefont
  {Eilertsen}, \citenamefont {J\"onsson}, \citenamefont {Ropinski},
  \citenamefont {Unger},\ and\ \citenamefont
  {Ynnerman}}]{Eilertsen2020_ClassifyingClassifier}%
  \BibitemOpen
  \bibfield  {author} {\bibinfo {author} {\bibfnamefont {G.}~\bibnamefont
  {Eilertsen}}, \bibinfo {author} {\bibfnamefont {D.}~\bibnamefont
  {J\"onsson}}, \bibinfo {author} {\bibfnamefont {T.}~\bibnamefont {Ropinski}},
  \bibinfo {author} {\bibfnamefont {J.}~\bibnamefont {Unger}},\ and\ \bibinfo
  {author} {\bibfnamefont {A.}~\bibnamefont {Ynnerman}},\ }\bibfield  {title}
  {\bibinfo {title} {Classifying the classifier: dissecting the weight space of
  neural networks},\ }\href {https://doi.org/10.48550/arXiv.2002.05688}
  {\bibfield  {journal} {\bibinfo  {journal} {arXiv:2002.05688}\ } (\bibinfo
  {year} {2020})}\BibitemShut {NoStop}%
\bibitem [{\citenamefont {Hernandes}\ \emph {et~al.}(2025)\citenamefont
  {Hernandes}, \citenamefont {Spriggs}, \citenamefont {Khaleefah},\ and\
  \citenamefont {Greplov\'{a}}}]{hernandes2025adiabatic}%
  \BibitemOpen
  \bibfield  {author} {\bibinfo {author} {\bibfnamefont {V.}~\bibnamefont
  {Hernandes}}, \bibinfo {author} {\bibfnamefont {T.}~\bibnamefont {Spriggs}},
  \bibinfo {author} {\bibfnamefont {S.}~\bibnamefont {Khaleefah}},\ and\
  \bibinfo {author} {\bibfnamefont {E.}~\bibnamefont {Greplov\'{a}}},\
  }\bibfield  {title} {\bibinfo {title} {Adiabatic fine-tuning of neural
  quantum states enables detection of phase transitions in weight space},\
  }\href {https://doi.org/10.48550/arXiv.2503.17140} {\bibfield  {journal}
  {\bibinfo  {journal} {arxiv:2503.17140}\ } (\bibinfo {year}
  {2025})}\BibitemShut {NoStop}%
\bibitem [{\citenamefont {Cybinski}\ \emph
  {et~al.}(2024{\natexlab{b}})\citenamefont {Cybinski}, \citenamefont {Enouen},
  \citenamefont {Georges},\ and\ \citenamefont {Dawid}}]{Cybinski2024Tetris}%
  \BibitemOpen
  \bibfield  {author} {\bibinfo {author} {\bibfnamefont {K.}~\bibnamefont
  {Cybinski}}, \bibinfo {author} {\bibfnamefont {J.}~\bibnamefont {Enouen}},
  \bibinfo {author} {\bibfnamefont {A.}~\bibnamefont {Georges}},\ and\ \bibinfo
  {author} {\bibfnamefont {A.}~\bibnamefont {Dawid}},\ }\bibfield  {title}
  {\bibinfo {title} {Speak so a physicist can understand you! {T}etris{CNN} for
  detecting phase transitions and order parameters},\ }\href
  {https://doi.org/10.48550/arXiv.2411.02237} {\bibfield  {journal} {\bibinfo
  {journal} {arxiv:2411.02237}\ } (\bibinfo {year}
  {2024}{\natexlab{b}})}\BibitemShut {NoStop}%
\bibitem [{\citenamefont {Liu}\ \emph {et~al.}(2025)\citenamefont {Liu},
  \citenamefont {Wang}, \citenamefont {Vaidya}, \citenamefont {Ruehle},
  \citenamefont {Halverson}, \citenamefont {Solja\v{c}i\'{c}}, \citenamefont
  {Hou},\ and\ \citenamefont {Tegmark}}]{liu2025kankolmogorovarnoldnetworks}%
  \BibitemOpen
  \bibfield  {author} {\bibinfo {author} {\bibfnamefont {Z.}~\bibnamefont
  {Liu}}, \bibinfo {author} {\bibfnamefont {Y.}~\bibnamefont {Wang}}, \bibinfo
  {author} {\bibfnamefont {S.}~\bibnamefont {Vaidya}}, \bibinfo {author}
  {\bibfnamefont {F.}~\bibnamefont {Ruehle}}, \bibinfo {author} {\bibfnamefont
  {J.}~\bibnamefont {Halverson}}, \bibinfo {author} {\bibfnamefont
  {M.}~\bibnamefont {Solja\v{c}i\'{c}}}, \bibinfo {author} {\bibfnamefont
  {T.~Y.}\ \bibnamefont {Hou}},\ and\ \bibinfo {author} {\bibfnamefont
  {M.}~\bibnamefont {Tegmark}},\ }\bibfield  {title} {\bibinfo {title} {{KAN}:
  {K}olmogorov-{A}rnold networks},\ }\href
  {https://doi.org/10.48550/arXiv.2404.19756} {\bibfield  {journal} {\bibinfo
  {journal} {arxiv:2404.19756}\ } (\bibinfo {year} {2025})}\BibitemShut
  {NoStop}%
\bibitem [{Sup({\natexlab{b}})}]{Supplement-code}%
  \BibitemOpen
  \href@noop {} {} ({\natexlab{b}}),\ \bibinfo {note} {the code for all
  simulations performed in the paper can be found at the following zenodo
  repository \doi{10.5281/zenodo.17102795} and gitlab:
  \url{https://gitlab.com/QMAI/papers/toponeuralnetworks}.}\BibitemShut {Stop}%
\bibitem [{Sta()}]{StanThesis}%
  \BibitemOpen
  \href@noop {} {}\bibinfo {note} {S. Bergkamp, ``Topological Phase Transition
  Learning. Unsupervised and Adversarial Methods'', MEP Thesis, TU Delft,
  (2024)}\BibitemShut {NoStop}%
\bibitem [{Bad()}]{BadrThesis}%
  \BibitemOpen
  \href@noop {} {}\bibinfo {note} {B. Zouggari, ``Simplifying Neural Networks
  for Quantum Classification'', BEP Thesis, TU Delft, (2024)}\BibitemShut
  {NoStop}%
\end{thebibliography}%
\let\oldaddcontentsline\addcontentsline
\renewcommand{\addcontentsline}[3]{}
\let\addcontentsline\oldaddcontentsline

\renewcommand\thesection{S\arabic{section}}
\renewcommand{\theHsection}{S\arabic{section}}
\renewcommand\thefigure{S\arabic{figure}}
\renewcommand{\theHfigure}{S\arabic{figure}}
\setcounter{figure}{0}
\setcounter{equation}{0}
\renewcommand\theequation{S\arabic{equation}}
\renewcommand{\theHequation}{S\arabic{equation}}
\newcommand{\be}{\begin{equation}}
	\newcommand{\ee}{\end{equation}}
\newcommand{\pih}{\frac{2\pi \mathcal{N}}{\hbar}}
\newcommand{\bea}{\begin{eqnarray}}
	\newcommand{\eea}{\end{eqnarray}}
\newcommand{\HH}{{\cal H}}
\newcommand{\RR}{{\cal R}}
\newcommand{\Li}{{\rm Li}}
\newcommand{\p}{\partial}
\newcommand{\s}{\sigma}
\newcommand{\la}{\left\langle}
\newcommand{\ra}{\right\rangle}
\newcommand{\lb}{\left[}
\newcommand{\rb}{\right]}
\newcommand{\lp}{\left(}
\newcommand{\rp}{\right)}
\newcommand{\sgn}{{\rm sgn\,}}
\renewcommand{\Re}{{\rm \, Re\,}}
\renewcommand{\Im}{{\rm \, Im\,}}
\newcommand{\sinc}{{\rm \, sinc\,}}
\renewcommand{\vec}[1]{{\boldsymbol #1}}
\renewcommand{\epsilon}{\varepsilon}
\newcommand{\rd}[1]{\mathop{\mathrm{d}#1}}
\newcommand{\tp}[1]{#1^\mathrm{T}}
\newcommand{\hc}[1]{#1^\dagger}
\def\nn{\nonumber\\}
\newcommand{\mpar}[1]{\marginpar{\small \it #1}}
\newcommand{\addLL}[1]{\textcolor{red}{#1}}
\newcommand{\addmove}[1]{\textcolor{green}{#1}}
\newcommand{\revision}[1]{\textcolor{red}{#1}}
\newcommand{\revisionA}[1]{\textcolor{blue}{#1}}
\newcommand{\revisionB}[1]{\textcolor{magenta}{#1}}
\newcommand{\revisionC}[1]{\textcolor{green}{#1}}
\setcounter{page}{1}
\newpage
\onecolumngrid
\begin{center}
	\textbf{Supplemental material for ``Why is topology hard to learn?''}\\
	by D. O. Oriekhov, Stan Bergkamp,  Guliuxin Jin,  Juan Daniel Torres Luna,  Badr Zouggari, Sibren van der Meer, Naoual El Yazidi, and Eliska Greplova
\end{center}
\twocolumngrid

\tableofcontents

\section{Derivation of real space winding number for the family of SSH models}

In this section, we demonstrate how the RSWN formula derived for the general AIII class topological insulator in Ref.\cite{mondragon2014topological} could be reduced to a unique form for the SSH-type family of insulators. The derivation includes the following steps: (i) definition of eigenbasis of chiral symmetry operator for SSH-type models with two sublattices; (ii) transformation of general RSWN expression to this basis by using cyclic properties of trace; (iii) derivation of restricting identities that follow from orthogonality condition of eigenstates with different energies; (iv) application of these identities to downfold the power of polynomials in RSWN formula from fourth order to second order; (v) presentation of additional restrictions from inversion symmetry on the lattice.

The RSWN for general AIII class topological insulator reads:
\begin{align}
W=-\frac{1}{L} \operatorname{Tr}\left\{P_B Q P_A\left[X, P_A Q P_B\right]\right\}.
\end{align}
It is expressed through the commutator of $Q$-operator, and projectors $P_{A,B}$ on two sets of sublattices $A,B$ conjugated by the $\Gamma=P_A-P_B$ operator, with the coordinate operator $X$. The operator $Q=P_{+}-P_{-}$ is composed of projectors on empty and filled bands, and the system length $L$ equals to number of unit cells $N$ times the lattice constant.
 
Due to the chiral symmetry $\Gamma$ that acts in sublattice space, the eigenstates have the following form:
\begin{align}
	&\Psi^{n,+}=\sum_{x}(\Psi^{n,+}_{x,A}|x,A\rangle+\Psi^{n,+}_{x,B}|x,B\rangle),\nonumber\\
    &\Psi^{n,-}=\sum_{x}(\Psi^{n,+}_{x,A}|x,A\rangle-\Psi^{n,+}_{x,B}|x,B\rangle).
\end{align}
Here we took into account that Hamiltonian has a block-antidiagonal form and $\Gamma \Psi^{n,+} = \Psi^{n,-}$ with $\Gamma=\sigma_3=\text{diag}(1,-1)$.
\begin{figure}
    \centering
    \includegraphics[scale=1]{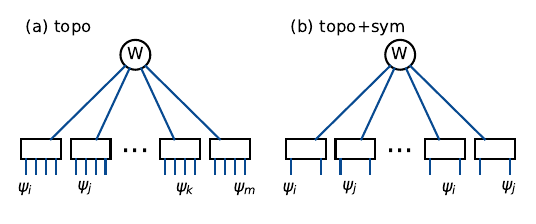}
    \caption{A change of architecture of a ``topology-inspired'' neural network under the properties of SSH-type modes at half-filling. The property of bands under chiral symmetry is included to change the tensor network layer from a 4-leg type to a 2-leg type. The corresponding sizes of input and the number of weights in the regression layer are (a) $N^4$ and (b) $N^2$.}
\label{fig-Appendix:conversion-4leg-to-2leg-input}
\end{figure}

The operators $Q_{AB/BA}$ have the form after substitution of positive and negative energy projectors, and cancellation of equal terms:
\begin{align}
	Q_{AB}=2\sum_x \sum_{x'} \sum_n |x,A\rangle \Psi^{n,+}_{x,A}\Psi^{n,+}_{x',B}\langle x',B|\\
	Q_{BA}=2\sum_x \sum_{x'} \sum_n |x,B\rangle \Psi^{n,+}_{x,B}\Psi^{n,+}_{x',A}\langle x',A|
\end{align}
Next, the winding number reads:
\begin{widetext}
\begin{align}
	W&=\frac{1}{L}\text{Tr}[Q_{BA}(\hat{X}Q_{AB}-Q_{AB}\hat{X})]=\frac{1}{L}\text{Tr}[Q_{BA}\hat{X}Q_{AB}]-\text{Tr}[Q_{AB}\hat{X}Q_{BA}]=\\
	&=\frac{4}{L}\text{Tr}[\sum_x \sum_{x'} \sum_n |x,B\rangle \Psi^{n,+}_{x,B}\Psi^{n,+}_{x',A}\langle x',A|  x'\sum_{x''} \sum_{n'} |x',A\rangle \Psi^{n',+}_{x',A}\Psi^{n',+}_{x'',B}\langle x'',B| ]-\nn
	&-\frac{4}{L}\text{Tr}[\sum_x \sum_{x'} \sum_n |x,A\rangle \Psi^{n,+}_{x,A}\Psi^{n,+}_{x',B}\langle x',B|  x'\sum_{x''} \sum_{n'} |x',B\rangle \Psi^{n',+}_{x',B}\Psi^{n',+}_{x'',A}\langle x'',A| ]
\end{align}
This formula could be converted into the expression presented in the main text, 
\begin{align}
    W=\frac{4}{L}\left(\sum_x \sum_{x'} \sum_n \sum_{n'}\Psi^{n,+}_{x,B}\Psi^{n,+}_{x',A}  x'  \Psi^{n',+}_{x',A}\Psi^{n',+}_{x,B} -\sum_x \sum_{x'} \sum_n \sum_{n'} \Psi^{n,+}_{x,A}\Psi^{n,+}_{x',B}  x'  \Psi^{n',+}_{x',B}\Psi^{n',+}_{x,A}\right).
\end{align}
\end{widetext}
From this expression the individual four-term products could be taken as inputs into `topo' neural network, see Fig.\ref{fig-Appendix:conversion-4leg-to-2leg-input}(a).

Due to only two sublattices, the following identities hold:
\begin{align}
	&\delta_{n,n'}=\langle\Psi^{n,+}|\Psi^{n',+}\rangle =\sum_{x}(\Psi_{x,A}^{n,+}\Psi_{x,A}^{n',+}+\Psi_{x,B}^{n,+}\Psi_{x,B}^{n',+} ),
	\nonumber\\
    &0=\langle\Psi^{n,+}|\Psi^{n',-}\rangle =\sum_{x}(\Psi_{x,A}^{n,+}\Psi_{x,A}^{n',+}-\Psi_{x,B}^{n,+}\Psi_{x,B}^{n',+} )
\end{align}
Taking sum and difference, we obtain the following results:
\begin{align}
	\sum_{x}\Psi_{x,A}^{n,+}\Psi_{x,A}^{n',+}=\sum_{x}\Psi_{x,B}^{n,+}\Psi_{x,B}^{n',+}=\frac{1}{2}\delta_{n,n'}
\end{align}
This result does not hold for localized edge states. Instead, such combinations give $1$ or $0$ (or a random float number in the vicinity of the phase transition). Thus, the corrected expression is:
\begin{align}	&\sum_{x}\Psi_{x,A}^{n,+}\Psi_{x,A}^{n',+}=\sum_{x}\Psi_{x,B}^{n,+}\Psi_{x,B}^{n',+}=\nonumber\\
&\begin{cases}
			\frac{1}{2}\delta_{n,n'} & n,n'\notin \{\text{edge state indices}\}\\
			\delta_{n,n'} \,\text{for A/B},\, 0\,\text{for B/A} & n=n'=\text{localized edge state}
		\end{cases}
\end{align}
Substitution of these relations into RSWN formula reduces isto the following sum of two-term products:
\begin{widetext}
\begin{align}
	W&=2 \sum_{n\notin\{\text{edge st.}\}} \sum_{x}x \left(\left[\Psi_{x, A}^{n,+}\right]^2-\left[\Psi_{x, B}^{n,+}\right]^2 \right)+\begin{cases}
		2 \sum_{n \in\{\text{hybr. edge states}\}} \sum_{x}x \left(\left[\Psi_{x, A}^{n,+}\right]^2-\left[\Psi_{x, B}^{n,+}\right]^2 \right), 
        \\
		4 \sum\limits_{n \in\{\text{loc. edge st., B}\}} \sum_{x}x \left[\Psi_{x, A}^{n,+}\right]^2-4 \sum\limits_{n \in\{\text{loc. edge st., A}\}} \sum_{x}x\left[\Psi_{x, B}^{n,+}\right]^2 
	\end{cases}
\end{align}
\end{widetext}

Such simplification significantly reduces the amount of redundant weights in `topo' NN, making their total amount to be $N^2$. The correspondig structure of the network is presented in Fig.\ref{fig-Appendix:conversion-4leg-to-2leg-input}(b).     However, this formula is still not unique for the SSH chain. Due to inversion symmetry, $\Psi_{x,A}^{n,+}=\Psi_{L-x,B}^{n,+}$, for each $n$ except the localized edge states. And the localized edge states themselves also transform into each other under a symmetry operation.  
E.g., the first part of the formula could be replaced by the following expressions: 
\begin{align}\label{eq:symmetry-rswn-ssh}
	W&=2 \sum_n \sum_{x}x \left(\left[\Psi_{L-x, B}^{n,+}\right]^2-\left[\Psi_{x, B}^{n,+}\right]^2 \right) =\nn
	&= 2 \sum_n \sum_{x}x \left(\left[\Psi_{x, A}^{n,+}\right]^2-\left[\Psi_{L-x, A}^{n,+}\right]^2 \right)=\nn
	&=2 \sum_n \sum_{x}x \left(\left[\Psi_{L-x, B}^{n,+}\right]^2-\left[\Psi_{L-x, A}^{n,+}\right]^2 \right).
\end{align}
Taking sums and differences of these expressions, we find that 
RSWN formula could be further restricted to 
\begin{align}
	W&=4 \sum_{n\notin{\text{edge states}}}^{L} \sum_{x>=0}^{L/2-1}x \left(\left[\Psi_{x, A}^{n,+}\right]^2-\left[\Psi_{x, B}^{n,+}\right]^2 \right).
\end{align}
The corresponding amount of inputs to NN is reduced to $N^2/2$ values.

\section{Classification by correlation of wave-function amplitudes with hopping parameter}
\label{appendix:classification-by-correlations-psi}
This section summarizes the results originally obtained in Ref. \cite{StanThesis} with the focus on why most neural networks do not automatically learn to capture topological aspects of condensed matter systems.

As was discussed in the main text, the FNN models take as input the wavefunction amplitudes and output the winding number or topological class. In this section, the inputs are composed of eigenstates of the SSH Hamiltonian, sorted by their respective energies. The label of the topological class is defined by the RSWN formula. The winding number in both finite and infinite systems is a monotonic function of hopping parameters ratio, $v/w$ \cite{mondragon2014topological,jin2024topologicalfinitesizeeffect}. Thus, if at least one amplitude is also a monotonic function of the $v/w$, the classification could be done with a single-perceptron neural network. As an example, neural network that has very high corresponding weight and bias placed exactly at the phase transition point would classify both trivial and topological classes instantly. 
\begin{figure*}
	\centering
	\includegraphics[scale=1]{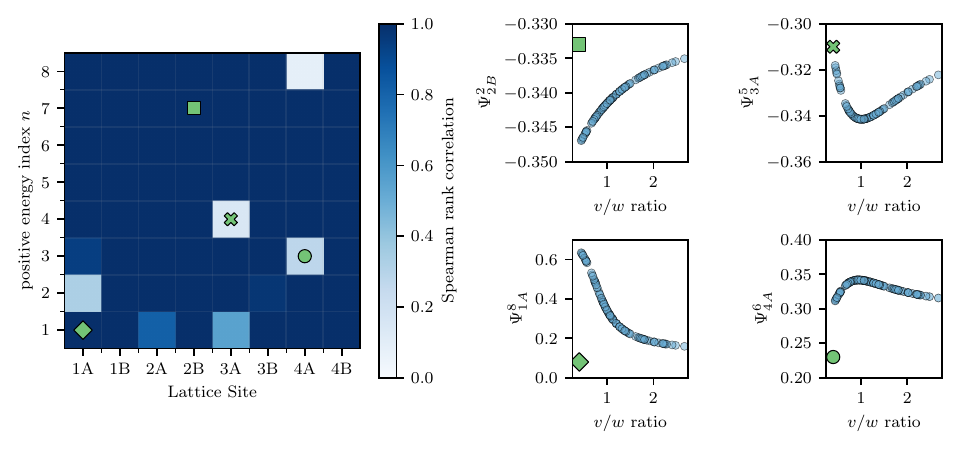}
	\caption{\textbf{Left}: A heatmap of the rank correlation coefficient at zero disorder is shown for individual components of SSH eigestates. Axes correspond to lattice site and energy index of eigenstate. The plot is restricted to positive energies and half of chain as the other indices correspond via particle-hole and inversion symmetries.  \textbf{Right}: Four panels marked by a green square, cross, diamong and circle, correspond to the similarly marked lattice sites in the left panel. Each panel shows the relationship of the amplitude values and $v/w$ for the respective wavefunction amplitude - lattice site combinations $\psi_{2,2B}, \psi_{5,3A}, \psi_{8,1A}, \psi_{6,4A}$.}
	\label{fig:sp_cor_distorted0}
\end{figure*}

To check the occurrence of such amplitudes within an input eigenstate, we perform numerical tests on SSH chain with $N=8$ unit cells. 

\subsection{Rank Correlation}
For the selected eigenstates we tested (shown in Fig.\ref{fig:sp_cor_distorted0}, right panel), we find multiple amplitudes to be a monotonic function of the $v/w$. This ratio is also the defining factor in the classification of the system as trivial or topological. Thus, any deep machine learning model could simply learn the ranking of the limited amount of amplitude data points and retrieve a threshold value for RSWN value crosses $0.5$ to identify the phase.

To test this numerical observation systematically, we determine the Spearman correlation rank of every amplitude of the calculated eigenstates. Positive correlations imply that as $v/w$ increases, so does wavefunction amplitude $\Psi^i$, and vice versa for negative correlations. If a threshold of the amplitude were used to classify the phase, then this rank correlation would be an upper limit of the accuracy. I.e., a rank correlation of 1 or -1 would mean perfect phase classification.

As can be seen in the left panel of Figure \ref{fig:sp_cor_distorted0}, almost all components have an absolute value of rank correlation equal to 1. In the four panels on the right, the wavefunction amplitude behavior is shown for some specific wavefunction/lattice site combinations ($\Psi^{2}_{2,B}, \Psi^{5}_{3,A}, \Psi^{8}_{1,A}, \Psi^{6}_{4,A}$). These panels show a smooth function of the amplitudes as a function of $v/w$. Two panels show perfectly monotonic behavior, while remaining two show non-monotonic behavior.

This analysis suggests that one of the reasons the FNN models may not learn the underlying topological patterns related to the calculation or estimation of topological information or invariants is that this data is easily classified. Therefore, the models do not need to extract these complex patterns to achieve high classification accuracy. The amplitudes are directly related to the hopping parameters and, therefore, the topological phases themselves. Using more complicated neural network architectures for this specific supervised learning task will therefore also yield the same results. 

This conclusion motivates us to study the performance of ML models discussed in the main text on disordered datasets as well. Disorder adds a `noise' into eigenstates images and breaks the possibility of classification with a single perceptron. This is discussed in detail below.

\subsection{Disorder in Hopping Parameters $v/w$}

To look at the effect of disorder of the individual hopping parameters on the amplitude data, the same rank correlation and corresponding amplitude scatter plots are given for a fixed relative value of disorder. The disorder is created by adding Gaussian zero-mean white noise (ZMWN). The relative disorder values indicate the width of the distribution (standard deviation) of the ZMWN that is added to each hopping parameter individually in the real-space Hamiltonian before diagonalization. In other words, all off-diagonal hopping elements in the Hamiltonian get a random noise term added.

\begin{figure*}
	\centering
	\includegraphics[scale=0.95]{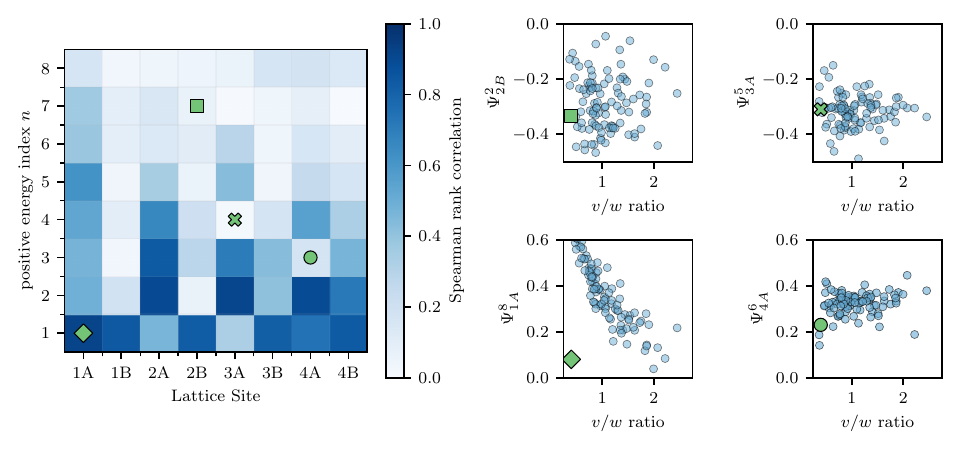}
	\caption{\textbf{Left}: A heatmap of the rank correlation coefficient for gaussian disorder with standard deviation $0.1$ is shown for individual components of SSH eigestates. Axes correspond to lattice site and energy index of eigenstate. \textbf{Right}: Four panels marked by a green square, cross, diamong and circle, correspond to the similarly marked lattice sites in the left panel. Each panel shows the relationship of the amplitude values and $v/w$ for the respective wavefunction amplitude - lattice site combinations $\psi_{2,2B}, \psi_{5,3A}, \psi_{8,1A}, \psi_{6,4A}$.}
	\label{fig:sp_cor_distorted2}
\end{figure*}

When ZMWN with a standard deviation of $0.10$ is added, the data becomes increasingly noisy, with the upper and lower energy states barely being correlated with labels (seen on the left of Fig.\ref{fig:sp_cor_distorted2}).  In the right panels of the Fig.\ref{fig:sp_cor_distorted2}, the four selected wavefunction amplitude - lattice site combinations increasingly show less correlation between the amplitudes and $v/w$. This indicates that the simple correlations between eigenstates and labels are removed from the dataset. Then, the FNN classifier trained on the clean dataset should demonstrate low performance on the disordered dataset - which appears in Figs.2 and 3 of the main text. 

\section{SVD analysis of FNN classifier}
This section contains results obtained during bachelor thesis project of B. Zouggari \cite{BadrThesis}.
In this section we analyze the correlations between individual eigenstates of SSH model and the topological phase. Based on these correlations, we construct a very simple feed-forward NNs that classify data with more that $90\%$ accuracy. We show that the SVD pruning could be used to build such networks from a standard classifier. The main drawback of such networks is that their performance drops dramatically on disordered systems (not unlike the example we described in the previous section).

\begin{figure}
	\centering
	\includegraphics[scale=0.9]{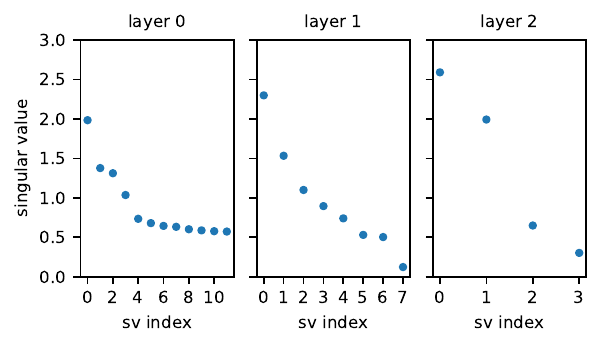}
	\caption{Singular values of the three layers for a FNN classifier. The architecture of the classifier for the $N=28$-unit cells ESSH chain is input $(2N)^2$, first hidden dimension $12$, second $8$ and output $4$. The accuracy after $20$ epochs of training was $99.7\%$ for the data set of size $\sim 50000$ with $25\%$ selected for validation.}
	\label{fig:Appendix-svd-values}
\end{figure}

\begin{figure}
	\centering
	\includegraphics[scale=1]{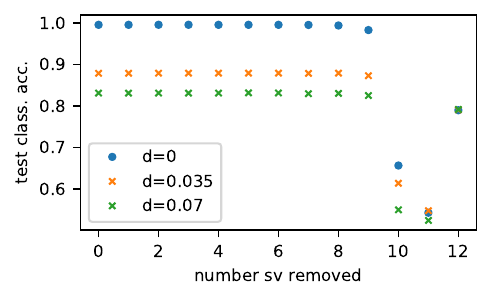}
	\caption{Accuracy of classification when several of the lowest singular values are replaced by zeros in the first layer weight matrix. The plot demonstrates that only the three to four largest singular values are responsible for the $99\%$ precision of classification. This suggests a possibility of a significant reduction of the total amount of weights present in the network.}
	\label{fig:Appendix-svd-accuracy}
\end{figure}

\begin{figure}
	\centering
	\includegraphics[scale=0.8]{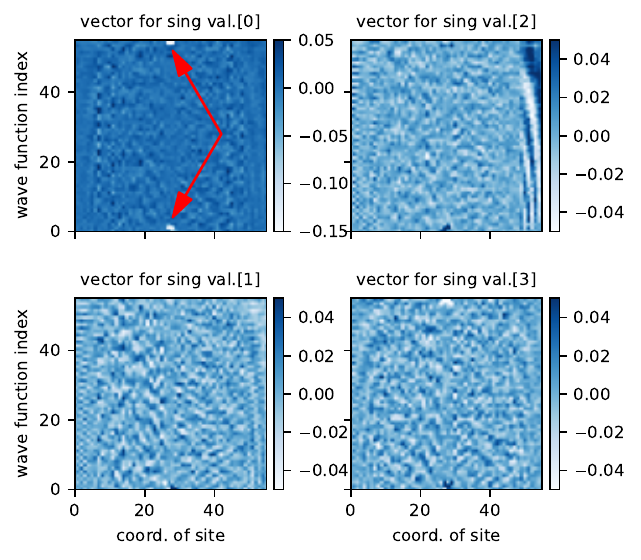}
	\caption{The four vectors of matrix $V$ in SVD decomposition of the first layer that correspond to the largest singular values. These four vectors are enough to achieve the $99\%$ accuracy of classification. The arrows show that these vectors are working like a feature maps that check for the existence of edge states. }
	\label{fig:Appendix-svd-featuremaps}
\end{figure}

The output $Y_i$ of the first layer of neural network is given by the following expression in terms of input $X_j$, weight matrix $\Delta_{ij}$ and bias $B_j$:
\begin{align}
	Y_i=Relu(\Delta_{ij} X_j + B_j)
\end{align}
The same form holds for the next layers as well.
The SVD applied to weight matrix reads
\begin{align}
	\Delta_{ij}=U_{ik}D_{kn}V_{nj},
\end{align} 
where $D_{kn}$ is a rectangular matrix with singular values on nonzero diagonal and both $U,V$ matrices are square matrices. This form of representation for weight matrix $\Delta$ suggests that it is possible to extract the linear feature-map transformations from the vectors within $V$. If we  remove a number of small singular values from $D$, we get a remaining set of corresponding relevant vectors in $U,V$. They form a projector operators:
\begin{align}
	\tilde{\Delta}_{ij}=\sum\limits_{d\in \{|D|>const\}} U_{id} D_{dd} V_{d j}.
\end{align}
The values of singular values within each layer for an example classifier with 3 layers are shown in Fig.\ref{fig:Appendix-svd-values}. The dependence of classification precision on singular values shown in Fig.\ref{fig:Appendix-svd-accuracy}.
The vectors $V_{d j}$ select significant superpositions of initial wave functions that play the main role in successful classification. Note that the orthogonality of eigenstates allows for inferring the direct importance of each state in such superposition. For example, it is reasonable to look only at edge nodes for zero-energy states. That is clearly shown in the density plots in Fig.\ref{fig:Appendix-svd-featuremaps}, as depicted by red arrows. An additional property of such a projection is that it performs a significant reduction of parameter space size between layers, playing the role of a large convolution kernel applied to the whole image. 

Based on Figs. \ref{fig:Appendix-svd-accuracy} and \ref{fig:Appendix-svd-featuremaps}, one concludes that the simplest FNN classifier, that maintains $98\%$ accuracy for ESSH phase classification, requires a hidden dimension of the second layer to be equal 2.  
 The corresponding pruned NN has just $N^2\times 2 + 2 \times 8 + 8 \times 4 $ parameters. This agrees well with the results in Sec.\ref{appendix:classification-by-correlations-psi} for the SSH model.
 
 Next, we show an intensity plot of individual `feature map' vectors that correspond to the largest singular values in Fig.\ref{fig:Appendix-svd-values}(a). The intensity plot shows that these vectors mainly check for the presence of localized edge states, as indicated by the arrows. In addition, they compare oscillatory patterns for particular eigenstates that are correlated the most with the topological phase label. This way of classification does not lead to learning topology. But, it still produces high accuracy as some states (in particular, the lowest energy states, which are transformed to edge states after phase transition) have a qualitative change in their structure. In Appendix.\ref{appendix:classification-by-correlations-psi} we show that for SSH model many of the states and their individual components are well correlated with the topological phase label. These components could be used to efficiently classify data with almost single-perceptron-type networks.

\section{Classification benchmark}
\label{sec:classification-benchmark}
As an additional benchmark, we test the performance of all NNs (topological, pre-processing and FNN) on the classification task for the ESSH model. Specifically, we will not try to regress on the value of the RSWN, we will simply try to predict in which of the four known classes the input wavefunction lies.
For ESSH model, there are in total 4 classes of winding numbers in finite-size geometry \cite{asboth2016short,HasanKane2010,su1979solitons,perez2018ssh,jin2024topologicalfinitesizeeffect}. The dataset contained 100k images of ESSH eigenstates without disorder. Each half of dataset was taken along different line in phase diagram of finite size chain of size $N=28$. The results show the `topo' network to perform much worse comparing to FNN classifier, while `preproc' network demonstrates the best performance. This could be explained by the fact that `topo' network is designated for perfect solution of regression task. To convert it into classifier, additional layer that converts outputs into classes was used. For precise architecture description see \cite{Supplement-code} and Sec.\ref{section-architect} below.
The extra amount of parameters in `preproc` network and its multilayer structure allowed 
for better learning and prediction of integer labels of classes.

\begin{figure}
    \centering
    \includegraphics[width=1.0\linewidth]{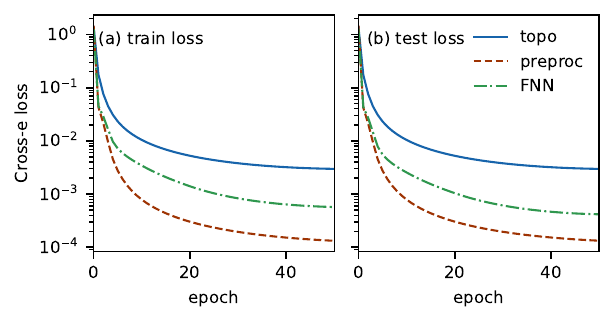}
    \caption{Benchmarking of three models (topological, preprocessed with middle layer size 12 and FNN with hidden layer sizes 12 and 8) on the one-hot ESSH classification task. Initialization is fully random for both models. Validation dataset size $0.15$, total dataset size 95k samples. All classes - RSWN $0, 1, 2, -1$ are present in equivalent proportions (e.g., same scale).}
    \label{fig:enter-label}
\end{figure}

\section{Neural network architectures}
\label{section-architect}
In this section we provide a technical description of the neural networks used in this study and the corresponding optimizers.

The loss functions that are optimized during training are: MSE loss for regression task, and Cross-entropy loss for classification task, both taken from pytorch.
Their exact expressions are:
\begin{align}
    &MSE =\sqrt{\sum\limits_{n=1}^{batch size} (W_n-W_{prediction, n})^2}\\
    &Cross-E=\ell(x, y)=L=\left\{l_1, \ldots, l_N\right\}^{\top}, \nonumber\\
     &l_n=-\sum_{c=1}^C w_c \log \frac{\exp \left(x_{n, c}\right)}{\sum_{i=1}^C \exp \left(x_{n, i}\right)} y_{n, c}.
\end{align}
ML models' architectures are summarized in Table \ref{tab:nns}.

\begin{table*}
    \centering
    \begin{tabular}{c|c|c|c|c|c|c}
        Model & task & Structure & TN layer & $\begin{array}{c}
        \text{Approximate number of}\\
        \text{trainable} \\ \text{parameters for} \\
        \text{N-cell system}
        \end{array}$ & Optimizer & general note \\
        \hline 
        `topo' & regression & $\begin{array}{c}
        \text{4-leg tensors} \\
        \text{dense layer}
        \end{array}$ & 4-leg inputs & $N^4$ & Adam, $lr=1e-4$ & $\begin{array}{c}
        \text{overparameterized}\\
        \text{for SSH}
        \end{array}$ \\
        \hline 
        `topo+sym' & regression & $\begin{array}{c}
        \text{2-leg tensors} \\
        \text{dense layer}
        \end{array}$ & 2-leg inputs & $N^2$ & Adam, $lr=1e-4$ & optimal for SSH \\
        \hline 
        `preproc+sym' & regression & $\begin{array}{c}
        \text{2-leg tensors} \\
        \text{dense layer} \\
        \text{RELu}\\
        \text{dense layer}
        \end{array}$ & 2-leg inputs & $N^2$ & $\begin{array}{c}
        \text{Adam, }lr=1e-4\\ \text{SGD, }lr=1e-2
        \end{array}
        $ & $\begin{array}{c}
        	\text{good optimization}\\
        	\text{possibilities} 
        \end{array}$  \\
        \hline  
        `FNN' & regression & $\begin{array}{c}
        \text{dense layer}\\
        \text{RELu}\\
        \text{dense layer}
        \text{RELu}\\
        \text{dense layer}
        \end{array}$ & None & $4 N^2$ & $\begin{array}{c}
        \text{Adam, }lr=1e-4\\ \text{SGD, }lr=1e-2
        \end{array}
        $ & $\begin{array}{c}
        	\text{overparameterized} \\
        	\text{for SSH}
        	\end{array}$ \\
        \hline 
        & & & & & & \\
        \hline 
        `topo+sym' & classification & $\begin{array}{c}
        \text{2-leg tensors} \\
        \text{dense layer}
        \end{array}$ & 2-leg inputs & $N^2$ & Adam, $lr=1e-4$ & optimal for SSH \\
        \hline 
        `preproc+sym' & classification & $\begin{array}{c}
        \text{2-leg tensors} \\
        \text{dense layer} \\
        \text{RELu}\\
        \text{dense layer}
        \end{array}$ & 2-leg inputs & $N^2$ & $\begin{array}{c}
        \text{Adam, }lr=1e-4\\ \text{SGD, }lr=1e-2
        \end{array}
        $ & $\begin{array}{c}
        	\text{good optimization}\\
        	\text{possibilities} 
        \end{array}$ \\
        \hline  
        `FNN' & classification& $\begin{array}{c}
        \text{dense layer}\\
        \text{RELu}\\
        \text{dense layer}
        \text{RELu}\\
        \text{dense layer}
        \end{array}$ & None & $4 N^2$ & $\begin{array}{c}
        \text{Adam, }lr=1e-4\\ \text{SGD, }lr=1e-2
        \end{array}
        $ & $\begin{array}{c}
        	\text{overparameterized} \\
        	\text{for SSH}
        \end{array}$ \\
        \hline
    \end{tabular}
    \caption{Models used in the study and their properties. The optmizers and learning rates are mentioned in the last column.}
    \label{tab:nns}
\end{table*}
\end{document}